\documentclass[pre,showpacs,superscriptaddress,nofootinbib,twocolumn]{revtex4}

\usepackage{graphicx}
\usepackage{amsmath}
\usepackage{amssymb}
\usepackage{amsfonts}
\usepackage{url}
\usepackage{color}
\usepackage{bm}
\usepackage[colorlinks=true,linkcolor=blue,citecolor=blue]{hyperref}
\usepackage[T1]{fontenc}

\def\Var{{\rm Var}}
\def\var{{\rm var}}

\def\n{\mathbf{n}}
\def\s{\mathbf{s}}
\def\x{\mathbf{x}}

\def\pa{\partial\Omega}

\def\P{{\mathbb P}}

\def\T{{\mathcal T}}
\def\L{{\mathcal L}}

\def\RH{{\mathcal R}}

\def\erf{\mathrm{erf}}
\def\erfc{\mathrm{erfc}}
\def\erfcx{\mathrm{erfcx}}

\begin{document}

\title{From single-particle stochastic kinetics to macroscopic reaction rates: \\
fastest first-passage time of $N$ random walkers}

\author{Denis~S.~Grebenkov}
\email{denis.grebenkov@polytechnique.edu}
\affiliation{Laboratoire de Physique de la Mati\`{e}re Condens\'{e}e (UMR 7643),
CNRS -- Ecole Polytechnique, IP Paris, 91128 Palaiseau, France}
\affiliation{Institute of Physics \& Astronomy, University of Potsdam,
14476 Potsdam-Golm, Germany}
\author{Ralf Metzler}
\email{rmetzler@uni-potsdam.de; corresponding author}
\affiliation{Institute of Physics \& Astronomy, University of Potsdam,
14476 Potsdam-Golm, Germany}
\author{Gleb Oshanin}
\affiliation{Sorbonne Universit\'e, CNRS, Laboratoire de Physique Th\'eorique de
la Mati\`ere Condens\'ee (UMR CNRS 7600), 4 Place Jussieu, F-75005, Paris, France}

\date{\today}

\begin{abstract}
We consider  the first-passage problem for $N$ identical
independent particles that are initially released uniformly in a
finite domain $\Omega$ and then diffuse toward a reactive area
$\Gamma$,
which can be part of the outer boundary of $\Omega$ or a reaction
centre in the interior of $\Omega$. For both cases of perfect
and partial reactions, we obtain the explicit formulas for the first
two moments of the fastest first-passage time (fFPT), i.e., the time
when the first out of the $N$ particles reacts with $\Gamma$.
Moreover, we investigate the full probability density of the fFPT.  We
discuss 
a significant role of the initial condition in the scaling of the
average fastest first-passage time with the particle number $N$,
namely, a much stronger dependence ($1/N$ and $1/N^2$ for partially and
perfectly reactive targets, respectively), in contrast to the well known
inverse-logarithmic behaviour found when all particles are released
from the same fixed point.
We combine analytic solutions with scaling arguments and stochastic
simulations to rationalise our results, which open new perspectives
for studying the relevance of multiple searchers in various situations
of molecular reactions, in particular, in living cells.
\end{abstract}





\maketitle

\section{Introduction}
\label{sec:intro}

Physical kinetics studies the dynamics of molecular chemical reactions in terms
of the time dependence of the reactant concentrations \cite{atkins}. If only
one reactant is present with concentration $[A]$ (unimolecular reaction),
the associated first-order rate equation is typically written in the
form $d[A]/dt= -k[A]$ with the reaction rate $k$. The resulting dynamics
is exponential, $[A]=[A]_0\exp(-kt)$, with initial concentration $[A]_0$
and half-life  time $t_{1/2}= \ln(2)/k$. A physical derivation of chemical
reaction rates in terms of the diffusivity of the molecular reactants was
given in the seminal 1916 paper by Smoluchowski \cite{smol} for immediate
coagulation. For instance, the Smoluchowski rate of a particle with diffusion
coefficient $D$ to hit an immobile spherical target of radius $a$ reads
$k_S=4\pi Da [A]_0$. A more general approach to calculate the reaction rate
combining reaction and diffusion limitation was formulated by Collins and
Kimball \cite{Collins49}. Different diffusion mechanisms can hereby lead
to an effective renormalisation of the target size in these theories. Thus,
when DNA-binding proteins search for their target site on a DNA chain, the
combination of three-dimensional diffusion with one-dimensional diffusion
along the DNA causes the target size $a$ to be replaced by an "effective
sliding length". In the "facilitated diffusion picture" this quantity is
based on the typical distance covered by the protein during sliding while
it intermittently binds to the DNA \cite{bvh,leonid,olivier,michael}. Such
predictions and their generalisations can by-now be measured routinely in
single-molecule assays \cite{gijs,austin,golding,xie,mark,elf,elf1}.

In chemical rate-based formulations of reaction kinetics such as those by
Smoluchowski or Collins and Kimball it is tacitly assumed that the reactants
are present at sufficiently high abundance, thus effecting smooth concentration
levels. Moreover, in a "well-stirred" reaction container spatial coordinates
can be neglected. By-now the understanding is that such theories provide
an adequate description of the chemical kinetics for most systems based on
the interplay of molecular reaction and diffusion---upon some refinements
and generalisations incorporating various physical factors missed in the
original works
\cite{calef,szabo,szabo2,weiss,Hanggi90,colloids,ralf,Benichou14,Oshanin,Grebenkov19b}.
Exceptions are provided by several particular reaction schemes in which, under
some rather restrictive constraints, so-called fluctuation-induced behaviour
emerges, see, e.g., \cite{bal,don,bur,ovch1,red,ovch2,bray,osh,osh2,tauber}
and references therein. One particular question concerns the
possible concentration-dependence of the effective reaction rates
\cite{calef,conc1,conc2,conc3,conc4}. Indeed, the original Smoluchowski
approach and many of its generalisations are only plausible for sufficiently
low albeit finite concentrations. At higher concentrations diffusive transport
as the rate-controlling factor becomes less important, and concurrently the
diffusion coefficients themselves acquire a dependence on the concentrations
of reactants and products.

Conversely, in a shift of general interest towards understanding the
kinetic behaviour of reactions in diverse biochemical and biophysical
systems low concentrations are increasingly considered to be a salient
feature. Indeed, many intracellular processes of signalling, regulation,
infection, immune reactions, and metabolism as well as transmitter release
in neurons occur upon the arrival of one or few biomolecules to small
specific regions \cite{alberts,snustad}. The well-stirredness assumption
based on simple diffusion models leads to the conclusion that DNA-binding
proteins rapidly homogenise in the cell. Experiments show, however, that
even in relatively small bacterial cells such transcription factors are
localised around their encoding gene and further inhomogeneity is caused by
the nucleoid state \cite{kuhlman}. Moreover, genes that are controlled by
a specific transcription factor tend to locate next to the gene encoding
this transcription factor \cite{kepes,kolesov}. This is consistent with
explicit models for intracellular gene regulation \cite{pulkkinen,prathit}.
Such gene regulatory signals often rely on nanomolar concentrations
\cite{mcadams,mcadams1}, similar to molecular concentration levels of
autoinducer molecules controlling the state of cell colonies in quorum sensing
\cite{kuttler,mugler,seno,trovato,oliver}. This causes strong fluctuations
of regulation events \cite{pulkkinen,oudenaarden,oberg,carmine}.

The nanomolar concentration range of relevant molecular species
renders the concept of reaction rates rather ill-defined. Due to the
lack of a sufficiently large number of molecules, reaction times
become strongly defocused and one cannot describe the system in
terms of a single time scale associated with a reaction, but rather
in terms of the full distribution of random times to a reaction event
\cite{thiago,aljaz,aljaz1,Grebenkov18o,Grebenkov18,GrebenkovFPT,Rupprecht15}.
Indeed, the shortest relevant time scale of the associated first-passage of
molecules to their reaction target in such situations is "geometry-controlled"
\cite{aljaz,Grebenkov18} in terms of "direct paths" from the molecules'
initial position to its target. This shortest characteristic time
scale is also the most probable in the reaction time distribution
\cite{Grebenkov18o,Grebenkov18}.  The longest time scale, typically orders of
magnitude longer than the most probable time scale, corresponds to "indirect"
trajectories that on their path hit the boundary of the confined volume and
thus lose any signature of their original position \cite{aljaz,Grebenkov18}.

Most approaches determining reaction rates or full probability densities
of reaction times rely on a single-particle scenario, yet, despite of
the small concentrations we alluded to above, typically a given number
of particles are searching for a common target in parallel, for instance,
several transcription factors seeking to bind to a specific binding site on
the cellular DNA. For particles searching for a common target in parallel two
relevant questions can be asked: (i) how much faster is the search process
when more than one searcher is present and (ii) which searcher comes first,
for instance, when we think of two different species of transcription factors
competing for the same binding site. We could also think of an entire colony
of cells, for instance, the many thousands of cells in a bacteria colony
\cite{biofilm,biofilm1}, competing with each other for which cell is able
to react first to a common environmental challenge. On a different level
such competitive scenarios come into play when sperm cells race towards the
egg cell. Indeed, such "particle number" effects on the reaction kinetics
of few-molecule diffusion-limited reactions have recently been addressed
\cite{Reynaud15,Basnayake19,holcman}. Generally, there is a range of systems
in molecular and cellular biology for which the number of species involved lies
in some intermediate range---much more than a few, but still much less than a
macroscopic number (i.e., of the order of Avogadro's number $N_A\approx 6\times
10^{23}$) \cite{Phillips}.  In particular, neuronal connections often occur
on a dendritic spine, where calcium is present in big amounts and arrival
of the fastest of the calcium ions can trigger a transduction \cite{fain}.
In another instance, the post-synaptic current is generated when the first
receptor of a neurotransmitter is activated, a process which is mediated by
the release and transport of several thousands of neurotransmitters from
the pre-synaptic terminal. In immunology, which hosts a variety of such
examples, a gene mechanism responsible for the selection and expression
of a specific membrane receptor on B-cells relies on many such receptors,
which bind directly to recognise a molecular unit of a pathogen.

A mathematical model in which $N$ Brownian particles start from the
same position simultaneously and search for an immobile small target
was first analysed by Weiss, Shuler, and Lindenberg almost 40 years
ago \cite{Weiss83}.  Various additional aspects of this problem were
subsequently considered
\cite{Abad12,Meerson15,Meerson15a,Ro17,Agranov18,Lawley19,Lawley19b,Lawley19c,Madrid20,mejia,mejia2,Bray13}.
Specifically it was already shown by Weiss and colleagues that the
mean first arrival time of the fastest of the $N$ Brownian searchers
to the target is inversely proportional to $\ln N$ as $N
\to\infty$. This means that the change of the "efficiency" of a
reaction is quite modest as compared to the quite large investment in
requiring a larger number of searchers. It was thus concluded in
\cite{Weiss83} that the speedup due to many searchers is a
comparatively minor effect: namely, to reduce the reaction time scale
in a noticeable way, one needs a very large amount of searchers. For
instance, to have a reduction by just a factor of 10, more than
$20,000$ searchers need to be deployed, an expensive number for many
biological scenarios. Indeed, it was recently argued in the context
of the tens of millions of sperm cells competing for a single egg in
higher mammals that when $N$ such searchers are launched from the same
initial location to find a given target, the decisive time scale is the
arrival time of the \emph{first\/} searcher, that is, the shortest time
\cite{Reynaud15,Basnayake19,holcman}.

In this paper, we address the reaction dynamics between an immobile
small target site and $N$ searchers diffusing inside a bounded domain
$\Omega$ of finite volume $|\Omega|$.  A major part of our analysis
pertains to a very general geometry of the system such that a bounded
domain may have an \textit{arbitrary} shape, with the only constraint
being that the boundary is \textit{smooth} and thus having a
finite area $|\partial \Omega|$.  The target $\Gamma$ with an area
$|\Gamma|$ and a characteristic extent $R$ can be placed either on the
boundary or in the bulk.  For illustrative purposes, we will use some
simple domains.

We take here a broader perspective beyond the mean shortest time
associated with the arrival of the fastest among $N$ Brownian
searchers. Namely, we analyse the full distribution function of the
time of the first reaction event. This comprises three different
relevant aspects.  First, as it is already known from the
single-searcher scenario that the full distribution of reaction times
spans several distinct characteristic time scales
\cite{Grebenkov18o,Grebenkov18,aljaz,GrebenkovFPT} ranging from the
above-mentioned most probable time, a crossover time scale from a
hump-like region to a plateau-like regime in which all values of the
first passage times are nearly equally probable, and ultimately, the
mean first passage time to the reaction event, followed by an
exponential decay. When $N$ searchers operate in parallel, we
establish the full probability density of the fastest first-passage
time.

Second, compared to previous works we include imperfect reactions
characterised by a finite intrinsic reaction constant $\kappa$ (with
$\kappa = \infty$ corresponding to a perfect reaction). Namely, for a
chemical reaction to be successful, it is not sufficient for the
diffusing molecule just to arrive to the reaction centre, but a
reaction activation barrier needs to be overcome
\cite{Collins49,Sano79,Sapoval94,Hanggi90,Grebenkov06,Grebenkov07a,Singer08,Bressloff08,Grebenkov10b,Grebenkov19,Grebenkov20}.
On top of this many biomolecules present a specific binding area,
further reducing the probability of immediate reaction on
encounter. As a consequence repeated collisions with the target are
required, leading to repeated excursions in the volume. The effected
further defocusing of the reaction times, analysed in detail for the
case $N=1$ in
\cite{Grebenkov18o,Grebenkov18,GrebenkovFPT}, was partially studied in
\cite{Lawley19b,Lawley19c} for searchers starting from the same
location.  Here we highlight some additional features.

The third and the most important difference to  most previous
works is that we consider the scenario in which the searching
particles initially are placed at \textit{distinct} random positions.
The characteristic properties are then obtained by averaging over
these fixed initial positions, which are supposed to be uniformly
spread across the confining domain.  Recall that for a target placed
away from a boundary, in the thermodynamic limit when both the number
of particles $N$ and the volume $|\Omega|$ of the confining domain
$\Omega$ tend to infinity while the concentration $N/|\Omega|$ is kept
finite (neither very small nor too large), the Smoluchowski approach
(see Appendix \ref{Smol}) provides an exact solution in the case of a
perfect reaction \cite{tachiya,blumen,sergei,agmon,searchjpa}. As
shown in \cite{Benichou00,Lawleynon} this even holds for the case of
imperfect reactions when the finite reactivity is modelled in terms of
a stochastic Poisson gating process. In this sense it can be argued
that our analysis provides a connection between single-molecule
reaction kinetics and standard chemical kinetics based on effective
reaction rates even in the case when targets are located on the
boundary.  We also stress the conceptual difference in dealing with
the uniform initial distribution as compared to a fixed starting
position for all searchers(see also \cite{Madrid20} for a
related discussion).  In both cases, the randomness of reaction times
follows from random realisations of the particle trajectories.
However, an additional source of randomness comes into play due to the
random initial placement of each particle within the confining
domain. This creates an intriguing new aspect to the problem, and we
analyse its impact on the reaction kinetics resorting to an analysis
of the typical behaviour, the averaged behaviour, and quantify
fluctuations around the averaged behaviour. 
A uniform initial condition leads to a drastically different behaviour
as compared to the case of a fixed starting point considered
previously in
\cite{Meerson15,Meerson15a,Lawley19,Lawley19b,Lawley19c}. 
In particular, as first noticed in \cite{Weiss83} and later explored
in \cite{Ro17,Agranov18,Madrid20}, in a \textit{finite} domain the
contribution to the effective rate due to a diffusive search for a
target decreases in proportion to $1/N^2$, while the contribution due
to a penetration through a barrier against reaction vanishes as $1/N$
in the limit $N \to \infty$. This means that (i) placing $N$ searchers
at distinct positions gives a substantial increase in the reaction
efficiency, which can be orders of magnitude larger as compared to the
case when all $N$ searchers start from the same point, (ii) as
compared to a standard Collins-Kimball relation, which is valid in the
thermodynamic limit and in which both contributions have the same
$1/N$-dependence on the number of searchers, here the contribution due
to a diffusive search acquires an additional power of a concentration
of searchers, (which is a strong concentration effect), and hence,
(iii) in the limit $N
\gg 1$ reactions become inevitably controlled by chemistry
(kinetically-controlled reactions) rather than by diffusion
(diffusion-controlled reactions).  Overall, the analysis developed
here provides a comprehensive insight into the binding kinetics of the
extremes of first-passage phenomena in finite systems with multiple
diffusing reactants competing for a reactive target.

The paper is organised as follows.  In Section \ref{sec:fixed} we discuss
the implications of fixed versus uniform starting positions. Section
\ref{sec:theory} contains our main theoretical results including a brief
overview of general properties for bounded domains (Section
\ref{sec:recall}), approximations for the mean first-reaction time
(\ref{sec:mean}) and the variance (\ref{sec:variance}), the long-time
behaviour of the volume-averaged probability density of the reaction
time (\ref{sec:long-time}), the fluctuations between individual
realisations (\ref{sec:fluctuations}), and the typical reaction time
density (\ref{sec:typical}). Section \ref{sec:discussion} is devoted
to the discussion of these results and their implications to chemical
physics and biological systems. Details of derivations and some
additional analyses are presented in the Appendices.

\section{Fixed versus randomly distributed starting point}
\label{sec:fixed}

Before proceeding to the results we discuss the uniform initial condition
analysed in this paper. In the conventional macroscopic description of
chemical kinetics, the reaction rate is computed by averaging the
probability density $\rho(t|\x_0)$ of the first-passage time to the
target from the starting point $\x_0$,
\begin{equation}
J(t)=\int\limits_\Omega d\x_0 \, c_0(\x_0) \, \rho(t|\x_0),
\end{equation}
where $c_0(\x_0)$ is the initial concentration profile of
particles. As we assume a uniform initial concentration in the bounded
confining domain $\Omega$ with finite volume $|\Omega|$, the reaction
rate is simply proportional to the volume-averaged probability
density, $J(t) \propto \overline{\rho(t)}$, with
\begin{equation}
\label{eq:rho_volume}
\overline{\rho(t)}=\frac{1}{|\Omega|}\int\limits_\Omega d\x_0\rho(t|\x_0).
\end{equation}
As a consequence, the reaction rate $J(t)$ incorporates two intertwined
sources of randomness: a random choice of the initial position
$\x_0$ and a random trajectory from $\x_0$ to the target. Even
though both affect the first-passage time distribution, their
respective roles are in fact not that well understood, thus deserving
a more specific investigation.

A volume-averaged description as entering equation
(\ref{eq:rho_volume}) is justified whenever the number of diffusing
particles is macroscopically large so that one can actually speak
about concentrations. In many of the biologically relevant settings
discussed above, however, the number of diffusing particles can be
small or moderately large (say, a few tens or a few thousand). One
may therefore question whether the above macroscopic approximation is
still applicable. In particular, what is the role of stochastic
fluctuations between different realisations of the starting points?
This question becomes particularly relevant in the short-time limit.
In fact, if there is a single searcher, the density $\rho(t|\x_0)$
decays exponentially fast at short times, as $e^{-c/t}$, where $c$ is
proportional to the squared distance to a target from $\x_0$. In
contrast, the volume average in equation (\ref{eq:rho_volume})
superimposes all $\rho(t|\x_0)$ including the contributions from those
particles that started infinitely close to the target are dominant. As
a result, $\overline{\rho(t)}$ exhibits not an exponential but a
power-law behaviour as $t\to 0$, as we detail below. One can conclude
that $\overline{\rho(t)}$ is not representative of an actual behaviour
here, as one could expect.  But what happens for $N$ particles, in
particular, in the large $N$ limit?  In other words, we aim at
uncovering the role of stochastic fluctuations related to random
initial positions of the particles as we change $N$.

\begin{figure}
\centering
\includegraphics[width=40mm]{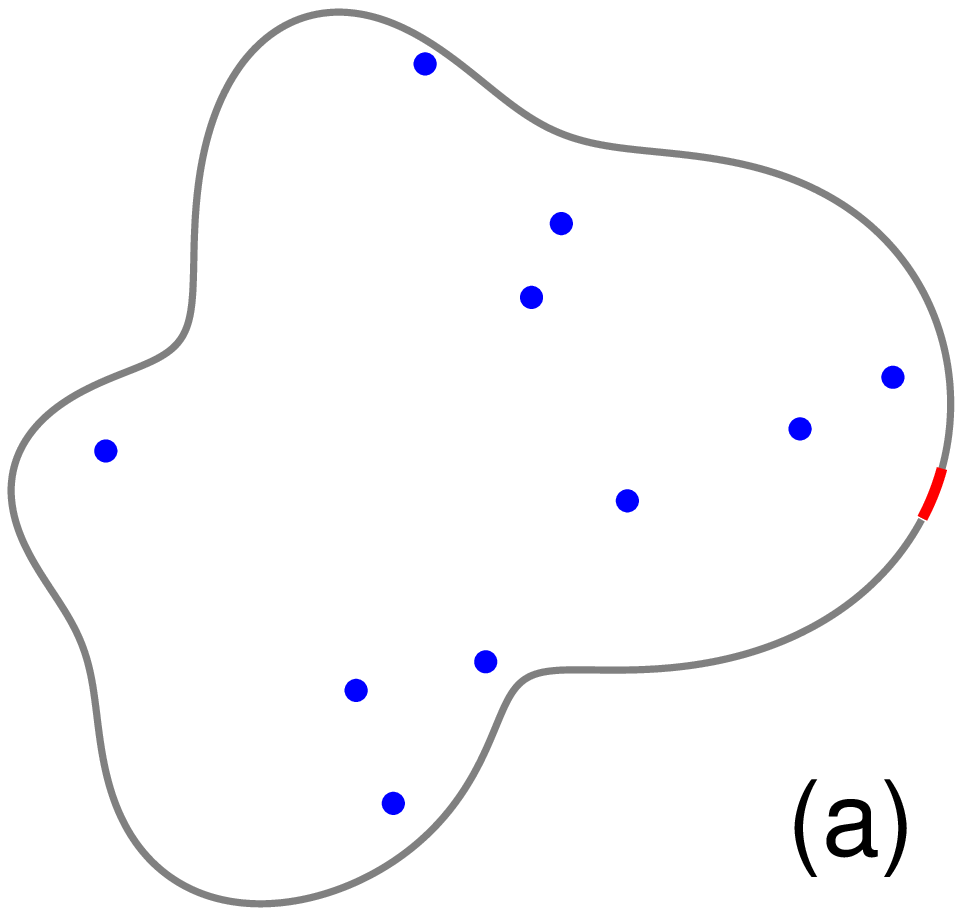} 
\includegraphics[width=40mm]{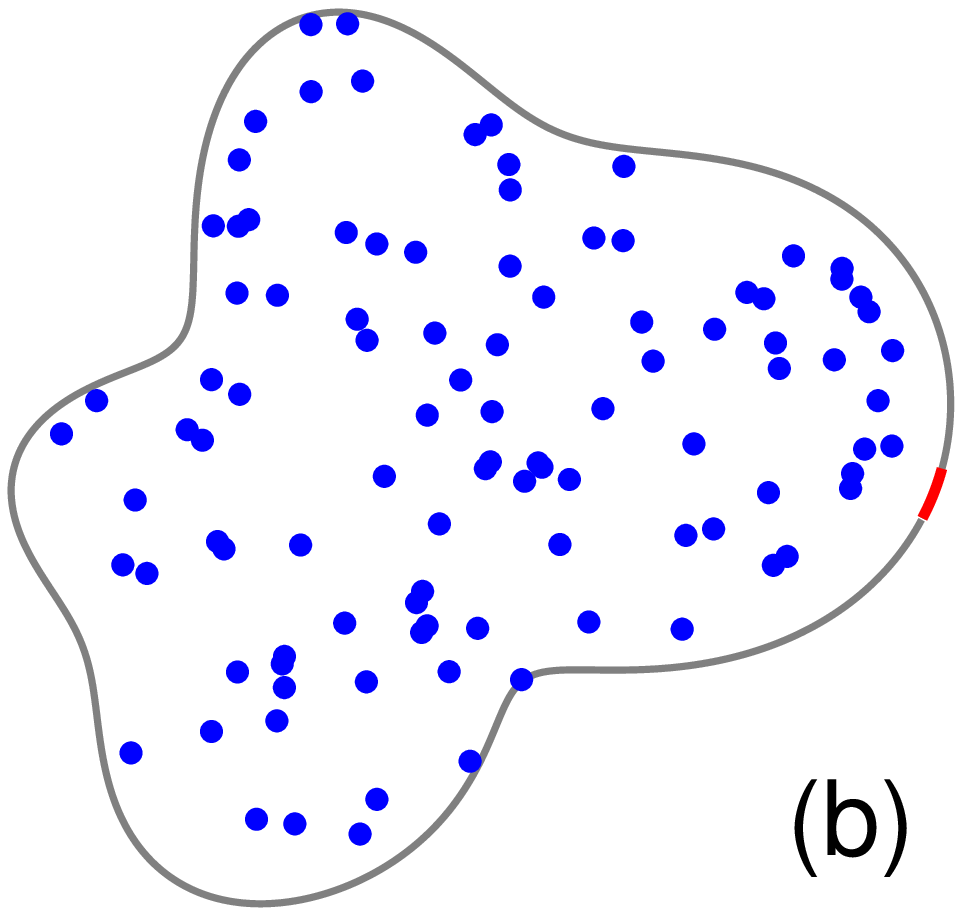} 
\caption{Arbitrary bounded domain $\Omega$ with a smooth reflecting boundary
$\partial\Omega$ (in grey) and a target (in red) located on the boundary
(similarly, the target can be located in the bulk). $N$ searchers (shown by
small blue circles) are initially randomly (uniformly) distributed
throughout the volume. {\bf (a)} $N = 10$, {\bf (b)} $N = 100$.}
\label{fig:domain}
\end{figure}

In what follows we investigate the gradual transition from the
single-particle description, in which stochastic fluctuations are
significant, to the macroscopic description. We consider $N$
independent identical particles undergoing Brownian dynamics with
diffusion coefficient $D$ that search in parallel for a common target
(Fig. \ref{fig:domain}). The first-passage time (FPT) $\tau_1$ for a
single particle, started from $\x_1$, is characterised by its survival
probability $S(t|\x_1) =\P_{\x_1}\{\tau_1>t\}$, from which the
probability density is obtained by differentiation,
\begin{equation}
\rho(t|\x_1)=-\frac{\partial}{\partial t}S(t|\x_1).
\end{equation}
The independence of the diffusing particles immediately implies that the
first-passage time ${\mathcal T}_N=\min\{\tau_1,\ldots,\tau_N\}$ of the
fastest particle among $N$, the so-called \textit{fastest} first-passage time
(fFPT), follows from the $N$-particle survival probability
\begin{equation}
\label{eq:SN_xx}
S_N(t|\x_1,\ldots,\x_N)=\P\{{\mathcal T}_N>t\}=\prod\limits_{n=1}^N S(t|\x_n).
\end{equation}
The knowledge of this survival probability for a single
particle therefore fully determines the one for multiple
non-interacting particles searching in parallel, and the problem of
characterising the fFPT may look trivial at first thought.

However, an exact explicit form of the survival probability $S(t|\x)$
is known for a very limited number of simple settings such as, for
instance, the FPT to the endpoints of an interval or to the boundary
of a sphere \cite{Redner}. In turn, only the asymptotic behaviour or
some approximate forms are known for most practically relevant cases
such as the narrow escape problem
\cite{Grigoriev02,Holcman04,Schuss07,Benichou08,Pillay10,Cheviakov10,Oshanin10,Cheviakov12,Grebenkov16,GrebenkovNEP}
(see also the review \cite{Holcman14}).  Moreover, even if $S(t|\x)$ is
known explicitly, finding the moments of the fFPT remains a
challenging and quite involved problem.  While most former studies of
these extreme first-passage times focused on the case when all
particles start from the same fixed point (i.e., $\x_1 = \ldots =
\x_N$) \cite{Weiss83,Basnayake19,holcman,Lawley19,Lawley19b,Lawley19c} we
here explore a different direction, namely, the case when the
particles start from independent and uniformly distributed points. In
other words, we aim at uncovering the role of stochastic fluctuations
related to random initial positions of the particles. Qualitatively,
as $N$ increases, the particular random realisation of the starting
points is expected to become irrelevant, and the macroscopic
description should become increasingly accurate. Here, we investigate
the transition from the single-particle setting to the macroscopic
limit and address the practically important question of when and how
such a description becomes applicable.

\section{Theory}
\label{sec:theory}

To develop the theoretical framework we consider a given realisation
of starting points $\x_1,\ldots,\x_N$. Then the probability density
function (PDF) of the fFPT to a target domain $\Gamma$ follows from
equation (\ref{eq:SN_xx}) in the form
\begin{eqnarray}
\nonumber
\rho_N(t|\x_1,\ldots,\x_N)&=&-\frac{\partial}{\partial t}[S(t|\x_1)\times
\ldots\times S(t|\x_N)]\\
&&\hspace*{-1.8cm}=S(t|\x_1)\times\ldots\times S(t|\x_N)\sum_{n=1}^N\frac{
\rho(t|\x_n)}{S(t|\x_n)}.
\end{eqnarray}
If $(\x_1,\ldots,\x_N)$ are independent and uniformly distributed
points in the bounded confining domain $\Omega$, the volume-averaged
density reads
\begin{eqnarray}
\nonumber
\overline{\rho_N(t)}&=&\int\limits_{\Omega}\frac{d\x_1}{|\Omega|}\ldots\int
\limits_{\Omega}\frac{d\x_N}{|\Omega|}\rho(t|\x_1,\ldots,\x_N)\\
&=&-\frac{d}{dt}\left[\overline{S(t)}\right]^N = N \left[\overline{S(t)}\right]^{N-1} \, \overline{\rho(t)} \,,
\label{eq:rho_N_vol}
\end{eqnarray}
where
\begin{equation}
\overline{S(t)}=\frac{1}{|\Omega|}\int\limits_\Omega d\x \,  S(t|\x)
\end{equation}
is the volume-averaged survival probability.

\subsection{Summary of general single-particle properties}
\label{sec:recall}

For a bounded domain $\Omega$ of an arbitrary shape with a smooth
boundary $\pa$ the survival probability admits in the most general case
the spectral expansion \cite{Redner,Gardiner,Risken}
\begin{equation}
\label{eq:S_x0}
S(t|\x_0)=\sum\limits_ne^{-Dt\lambda_n}u_n(\x_0)\int\limits_{\Omega}d\x \, u_n^*(\x),
\end{equation}
where the asterisk denotes the complex conjugate, $\lambda_n$ are
non-negative eigenvalues which are sorted in ascending order,
\begin{equation*}
0\leq\lambda_1\leq\lambda_2\leq\ldots\leq\lambda_n\leq\ldots, 
\end{equation*}
and the $u_n(\x)$ are $L_2(\Omega)$-normalised eigenfunctions of the Laplace
operator, $-\Delta u_n=\lambda_nu_n$, subject to appropriate boundary
conditions.  The eigenvalues  and eigenfunctions encode all the
information about the geometry of the domain, the location of the target and
its size \cite{Grebenkov13}.  We consider three typical situations (in
increasing order of generality): \\
(i) a fully reactive boundary described by the Dirichlet boundary
condition $(u_n)|_{\pa}=0$  (here, $\Gamma = \pa$); \\
(ii) a partially reactive boundary described by the Robin boundary
condition $\bigl(D\partial_{\n}u_n+ \kappa u_n)_{|\pa}=0$, where
$\kappa$ is the reactivity and $\partial_{\n}$ is the normal
derivative oriented outwards the domain (here, $\Gamma = \pa$); and \\
(iii) a partially reactive target $\Gamma$ on the otherwise reflecting
boundary $\pa\backslash \Gamma$, described by the mixed boundary
condition $\bigl(D\partial_{\n}u_n+ \kappa u_n)_{|\Gamma}=0$ and
$(\partial_{\n} u_n)_{|\pa\backslash\Gamma}=0$.  We emphasise
that the last situation includes two distinct cases: a target located
on the reflecting boundary and a target located in the bulk
(Fig. \ref{fig:domain}).  In the latter case, the boundary of the
confining domain $\Omega$ has two disjoint components: the outer
reflecting boundary and the inner reactive target.  Even though these
two locations of the target are usually distinguished in physical
literature, their mathematical description is the same.  Note that the
target region $\Gamma$ does not need to be connected, i.e., one can
consider multiple targets.

The volume-averaged survival probability then reads
\begin{equation}
\label{eq:S_vol_def}
\overline{S(t)}=\sum\limits_nc_ne^{-D\lambda_n t},
\end{equation}
where
\begin{equation}
\label{eq:cn}
c_n=\frac{1}{|\Omega|}\left|\int\limits_{\Omega}d\x \, u_n(\x)\right|^2.
\end{equation}
Similar spectral expansions hold for the PDF and its volume average,
\begin{equation}
\label{eq:rho_x0}
\rho(t|\x_0)=\sum\limits_n D\lambda_ne^{-Dt\lambda_n}u_n(\x_0)\int\limits_{
\Omega}d\x \, u_n^*(\x)
\end{equation}
and 
\begin{equation}
\overline{\rho(t)}=\sum\limits_nD\lambda_nc_ne^{-D\lambda_nt}.
\end{equation}
From these expansions one can easily derive the volume-averaged
moments of the FPT $\T_1$,
\begin{equation}
\label{eq:T1k}
\overline{\langle\T_1^k\rangle}=k!\sum\limits_n\frac{c_n}{(D\lambda_n)^k}\qquad
(k = 1,2,3,\ldots).
\end{equation}
Throughout the paper we distinguish the volume average over random initial points
(denoted by the overline) and the ensemble average over random trajectories
(denoted by angular brackets).

At long times the survival probability and the PDF decay exponentially fast,
with decay rate $1/(D\lambda_1)$ determined by the smallest eigenvalue
$\lambda_1$. The volume average does not affect this decay. In contrast, the
short-time behaviour is drastically different for fixed-point and volume-averaged
quantities. In fact, the volume-averaged survival probability behaves as
\begin{subequations}   \label{eq:Svol_short}
\begin{eqnarray}  \label{eq:Svol_short_perfect}
\overline{S(t)} &\simeq& 1-\frac{2|\Gamma|}{\sqrt{\pi}|\Omega|} \, \sqrt{Dt}+O(t) \quad (\kappa=\infty),\\  \label{eq:Svol_short_partial}
\overline{S(t)} &\simeq& 1-\frac{|\Gamma| \kappa }{|\Omega|} \, t + O(t^{3/2}) \hskip 8.5mm (\kappa<\infty),
\end{eqnarray}
\end{subequations}
which follows from the short-time expansion of the heat content
\cite{vandenBerg89,vandenBerg94,Desjardins94,Gilkey} (see also
Appendix \ref{sec:mfFPT} for the next-order correction).  Qualitative
arguments behind this asymptotic result are simple. In the
diffusion-limited regime ($\kappa=\infty$) only those particles in a
thin layer of width $\sqrt{Dt}$ near the reactive boundary $\Gamma$
can reach this boundary and thus disappear within short time. The
relative fraction of such particles is $\sqrt{Dt}\, |\Gamma|/|\Omega|$,
where $|\Gamma|$ is the surface area of $\Gamma$ and $|\Omega|$ is the
volume of the domain $\Omega$. In contrast, in the reaction-limited
regime, the decay of $\overline{S(t)}$ is limited by $\kappa$ and is
thus proportional to $t$.  An immediate consequence of
Eq. (\ref{eq:Svol_short}) is
\begin{equation}
\label{eq:rhovol_short}
\overline{\rho(t)}\simeq\frac{|\Gamma|}{|\Omega|}\times\left\{\begin{array}{ll}
\dfrac{1}{\sqrt{\pi}}\sqrt{D/t}+O(1),&\kappa=\infty,\\
\kappa+O(\sqrt{t}),&\kappa<\infty.\\\end{array}\right.
\end{equation}
In contrast, $S(t|\x_0)$ and $\rho(t|\x_0)$ exhibit a fast exponential
decay as $t\to 0$,
\begin{subequations}
\label{eq:Srho_short_x0}
\begin{eqnarray}
\label{eq:S_short_x0}
1-S(t|\x_0)\simeq At^{-\alpha}e^{-|\x_0-\pa|^2/(4Dt)},\\
\rho(t|\x_0)\simeq\alpha At^{-\alpha-1}e^{-|\x_0-\pa|^2/(4Dt)},
\label{eq:rho_short_x0}
\end{eqnarray}
\end{subequations}
with the constants $A$ and $\alpha$, see
\cite{Varadhan67,Grebenkov10a,aljaz,Grebenkov17d,Basnayake19,Lawley19c}.

As shown in \cite{Madrid20}, the above short-time asymptotic behaviour
of the survival probability for a single particle determines the
leading behaviour of the moments of the fFPT and its volume-averaged
probability density $\overline{\rho_N(t)}$ in the limit $N\to \infty$.
In particular, for a uniform distribution of the starting points,
$\overline{\rho_N(t)}$ is close to the Weibull density
\begin{equation}
\label{eq:Weibull}
\overline{\rho_N(t)}\simeq kBNt^{k-1}e^{-BNt^k}\quad(N\to\infty),
\end{equation}
where $k=1/2$ and $B=2|\Gamma|\sqrt{D}/[\sqrt{\pi}|\Omega|]$ for a perfectly
reactive target, while $k=1$ and $B=|\Gamma|\kappa/|\Omega|$ for a partially
reactive target.

\subsection{Volume-averaged mean fFPT}
\label{sec:mean}

We now turn to the many-particle case and obtain a simple approximation for
the volume-averaged mean fFPT
\begin{equation}
\label{eq:MfFPT_1d_volume}
\overline{\langle\T_N\rangle}=\int\limits_0^\infty dt \, t \, \overline{\rho_N(t|\x_1,
\ldots,\x_N)}=\int\limits_0^\infty dt \, \left[\overline{S(t)}\right]^N,
\end{equation}
where we used equation (\ref{eq:rho_N_vol}).  We first split this
integral into two parts,
\begin{equation}  \label{eq:MfFPT_auxil}
\overline{\langle\T_N\rangle}=\int\limits_0^T dt \, \left[\overline{S(t)}\right]^N + 
\int\limits_T ^\infty dt \, \left[\overline{S(t)}\right]^N.
\end{equation}
The first integral will be evaluated explicitly by using the
short-time asymptotic formula (\ref{eq:Svol_short}). Under an
appropriate choice of $T$, the contribution of the second term is
greatly attenuated for large $N$ and can be ignored.  The "threshold
time" $T$ can be chosen to minimise the error of such an approximation
but we will see that the final result does not depend on $T$, as
expected.

\subsubsection{Perfect reactivity}

\begin{figure}
\centering
\includegraphics[width=88mm]{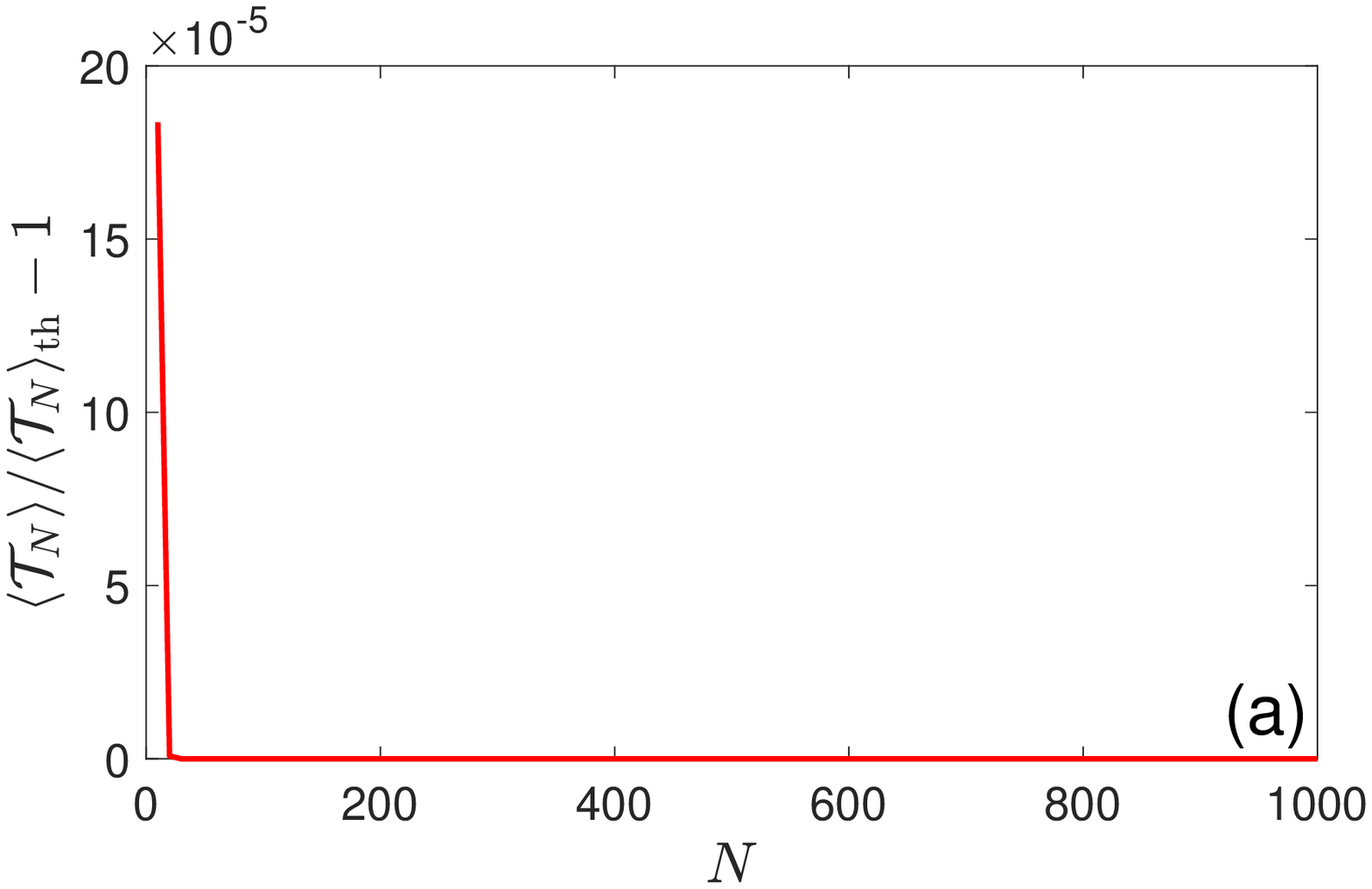} 
\includegraphics[width=88mm]{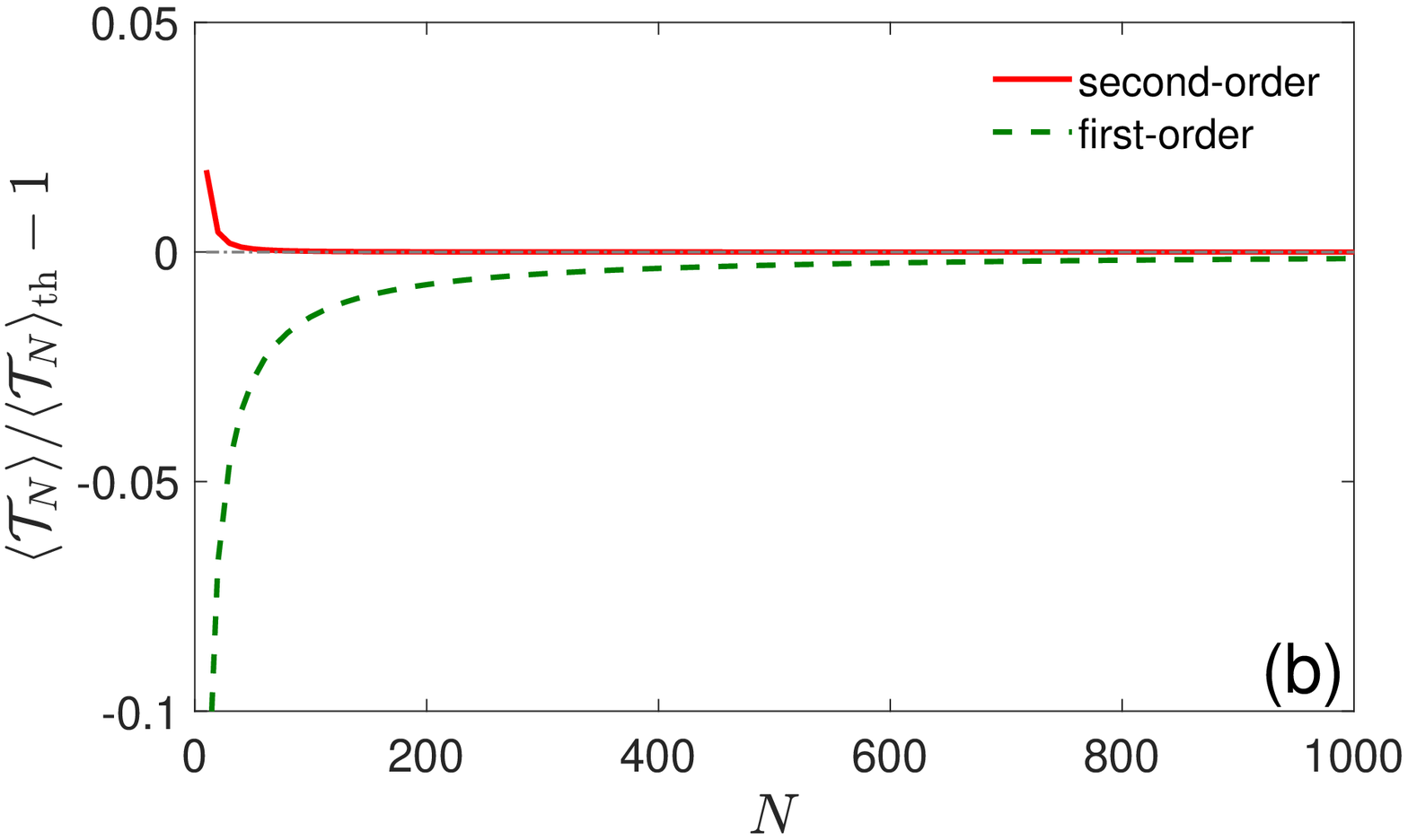} 
\includegraphics[width=88mm]{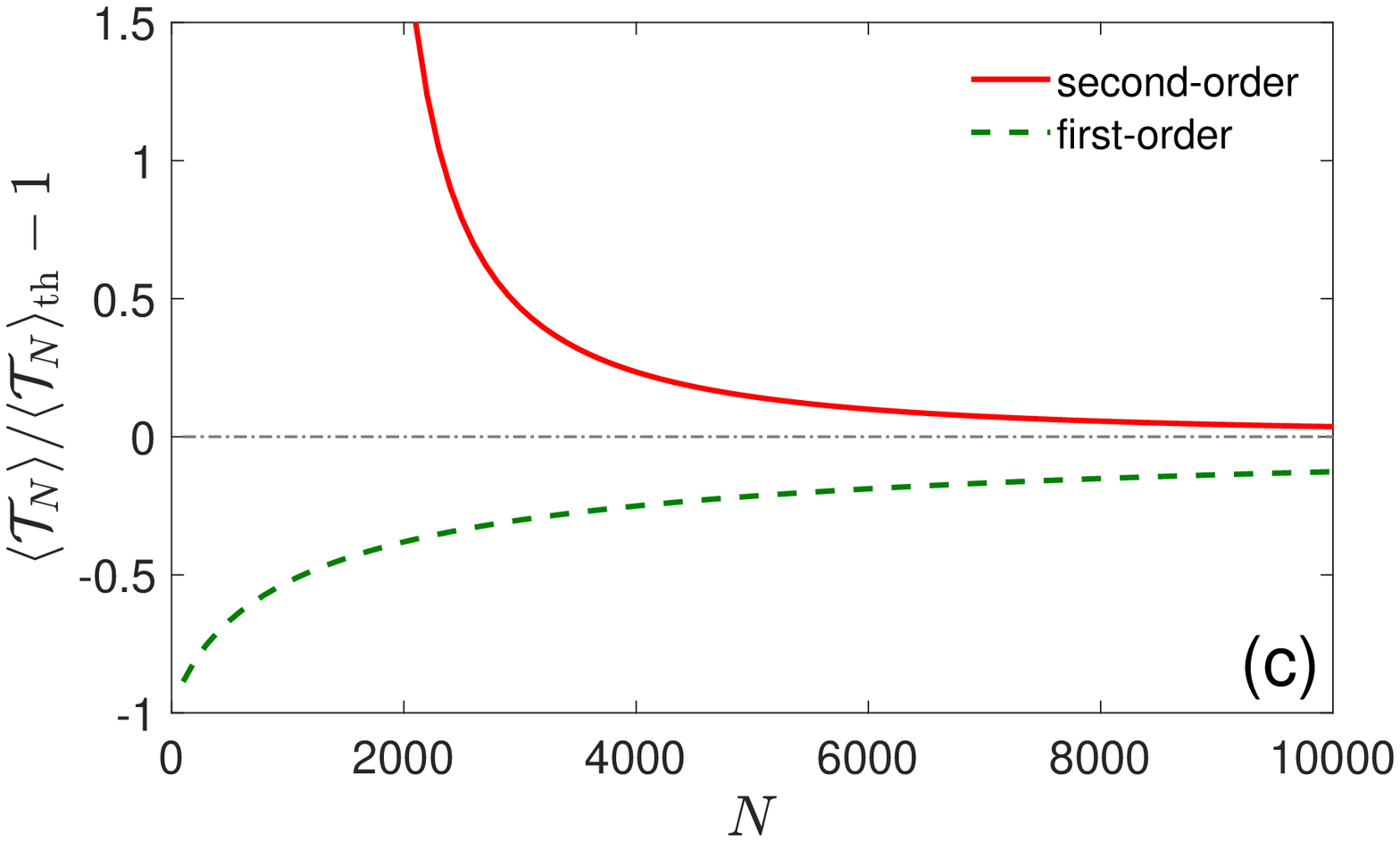} 
\caption{Relative error of our approximation of the volume-averaged mean
fFPT for: {\bf (a)} an interval $(0,L)$ with absorbing endpoints, {\bf
(b)} an absorbing sphere of radius $L$, and {\bf (c)} an absorbing
spherical target of radius $R = 0.1 L$ surrounded by a reflecting
sphere of radius $L$, with $L=1$ and $D=1$. The solid line shows the
relative error between $\overline{\langle\T_N \rangle}$ computed by
integrating numerically equation (\ref{eq:MfFPT_1d_volume}) and from
our approximation (\ref{eq:MfFPT_1d_approx}) for the interval, and
(\ref{eq:MfFPT_1d_approx_new}) for spheres. The dashed line indicates the
relative error for the case when only the leading term is kept in
Eq. (\ref{eq:MfFPT_1d_approx_new}).  For the interval the
approximation is remarkably accurate because the next-order correction
to the survival probability is exponentially small.  For a small
absorbing target, the asymptotic regime is established at much larger
$N$.  Note that $\overline{S(t)}$ was computed here via the spectral
expansion (\ref{eq:Sbar_target}) truncated after 10,000 terms to ensure
accurate integration at short times.}
\label{fig:rho1d_uniform_mean}
\end{figure}

For a perfectly reactive target ($\kappa=\infty$) and a finite
domain we use Eq. (\ref{eq:Svol_short_perfect}) to
get
\begin{equation}
\overline{\langle\T_N\rangle}\simeq\frac{\pi|\Omega|^2}{2D|\Gamma|^2}\frac{1-(1+
z(N+1))(1-z)^{N+1}}{(N+1)(N+2)},
\end{equation}
where $z=\frac{2|\Gamma|\sqrt{DT}}{\sqrt{\pi}|\Omega|}$ is a number
between $0$ and $1$ that is set by our choice of $T$. When
\begin{equation}  \label{eq:cond_N}
N|\ln(1-z)|\gg1 
\end{equation}
the correction term $(1-z)^{N+1}$ can be neglected, and one gets
\begin{equation}
\label{eq:MfFPT_1d_approx}
\overline{\langle\T_N\rangle}\simeq\frac{\pi|\Omega|^2}{2D|\Gamma|^2}\frac{1}{(N+1)
(N+2)}\qquad(N\gg1).
\end{equation}
This approximate expression indeed does not depend on $T$, as it
should.  Figure \ref{fig:rho1d_uniform_mean}(a) illustrates the
remarkable accuracy of this approximation in the case of an interval
$(0,L)$ with absorbing endpoints (here, $|\Gamma|/|\Omega|=2/L$). This
high accuracy results from the fact that the asymptotic relation
(\ref{eq:Svol_short}) is exponentially accurate for the interval,
\begin{equation}
\overline{S(t)} \simeq 1-\frac{4\sqrt{Dt}}{L\sqrt{\pi}}+O\left(e^{-L^2/(4Dt)}
\right)\quad (t\to 0),
\end{equation}
see the exact representation (\ref{eq:S1d_volume}).  For this reason,
we kept the form $1/[(N+1)(N+2)]$ in
Eq. (\ref{eq:MfFPT_1d_approx}) because its replacement by the leading
term $1/N^2$ would yield the correction $O(N^{-3})$.

In general, however, the next-order term in the short-time
behaviour of the survival probability is of order $O(t)$ that
deteriorates the exponential accuracy of
Eq. (\ref{eq:MfFPT_1d_approx}).
In Appendix \ref{sec:mfFPT}, we compute the correction $O(N^{-3})$
term to this formula:
\begin{equation}
\label{eq:MfFPT_1d_approx_new}
\overline{\langle\T_N\rangle} \simeq \frac{\pi|\Omega|^2}{2D|\Gamma|^2}\frac{1}{N^2} 
\biggl(1 - \biggl(1 - \frac{\pi |\Omega|}{2|\Gamma| \RH}\biggr) \frac{3}{N} + O(N^{-2})\biggr) ,
\end{equation}
where $\RH$ is the mean curvature radius of the target defined in
Eq. (\ref{eq:R_def}) (note that $\RH$ can be negative, see below).
The leading, $1/N^2$-term was first mentioned in the seminal
paper \cite{Weiss83} and later re-discovered in \cite{Ro17,Agranov18}.
Moreover, a transition from an intermediate $1/N$ scaling to the
ultimate $1/N^2$ behaviour was analysed in \cite{Ro17}.  A
mathematical derivation of the leading $1/N^2$-term for a rather
general class of diffusion processes was recently reported in
\cite{Madrid20}.
Figure \ref{fig:rho1d_uniform_mean}(b) shows the relative error of our
approximation for diffusion inside a ball of radius $L$ toward its
absorbing boundary (i.e, $\Gamma = \pa$ and $\RH = L$).  As expected,
the accuracy is lower at small $N$ but still remains excellent.
Finally, Fig. \ref{fig:rho1d_uniform_mean}(c) presents the relative
error for the case of a perfectly absorbing spherical target of radius
$R = 0.1 L$ surrounded by a reflecting sphere of radius $L$ (here, the
mean curvature radius is negative: $\RH = - R$).  In contrast to
previous examples, the asymptotic regime is established at much larger
$N$, in agreement with the condition (\ref{eq:cond_N}).  In fact, here
$z \propto |\Gamma| \propto (R/L)^2$ is 100 times smaller than in the
case of an absorbing sphere, i.e., one needs a hundred-fold increase
of $N$ to get a comparable accuracy.

How can one rationalise the unexpected $1/N^2$ scaling? As the starting points
of the searchers are spread in the domain, the closest particle to the target
has more chances to reach it first. As discussed in Appendix \ref{sec:closest},
the average distance $\overline{\delta}$ between the closest particle and the
target is of the order of $1/N$ (and not $1/N^{1/d}$ as one might expect from
a standard estimate of the average inter-particle distance in $d$ dimensions).
Moreover, if $N \gg 1$, there are many particles that start from a comparable
distance. Even if the closest particle will diffuse far away from the target,
there is a high chance that one of the other particle started at the distance
of order $1/N$ will hit the target. In other words, this problem resembles
the diffusion of a single particle on the interval $(0, \overline{\delta})$
with absorption at $0$ (on the boundary) and reflection on $\overline{\delta}$,
for which the mean FPT is $\overline{\delta}^2/(2D) \propto 1/N^2$. We conclude
that the scaling $1/N^2$ arises from the fact that many particle can start
close to the target. Moreover, the generic properties of the heat kernel
ensure that this scaling still holds for non-uniform initial distributions
which do not exclude particles from a close vicinity of the target (see
Appendix \ref{sec:closest}). As we will discuss in Sec. \ref{sec:discussion},
the decay $\overline{\langle\T_N\rangle} \propto N^{-2}$ is much faster and
drastically different from the scaling law $\langle\T_N\rangle \propto 1/\ln
N$ obtained earlier in the case of a fixed starting point \cite{Weiss83}.

\subsubsection{Partial reactivity}

For a partially reactive target ($\kappa < \infty$),
Eq. (\ref{eq:Svol_short_partial}) entails a different scaling for the
volume-averaged mean fFPT with $N$,
\begin{equation}
\label{eq:MfFPT_1d_approx2}
\overline{\langle\T_N\rangle}\simeq\frac{|\Omega|}{\kappa|\Gamma|} N^{-1} 
\biggl(1 + \biggl(\frac{\kappa |\Omega|}{D |\Gamma|}\biggr)^{\frac{1}{2}}N^{-\frac{1}{2}}
+O(N^{-1})\biggr),
\end{equation}
i.e. the contribution to the volume-averaged mean fFPT in the regime
of a reaction control vanishes only as a first inverse power of $N$.
The derivation of the correction term is presented in Appendix
\ref{sec:mfFPT}.  The leading $1/N$-term was recently obtained
in \cite{Madrid20}.

\begin{figure}
\centering
\includegraphics[width=88mm]{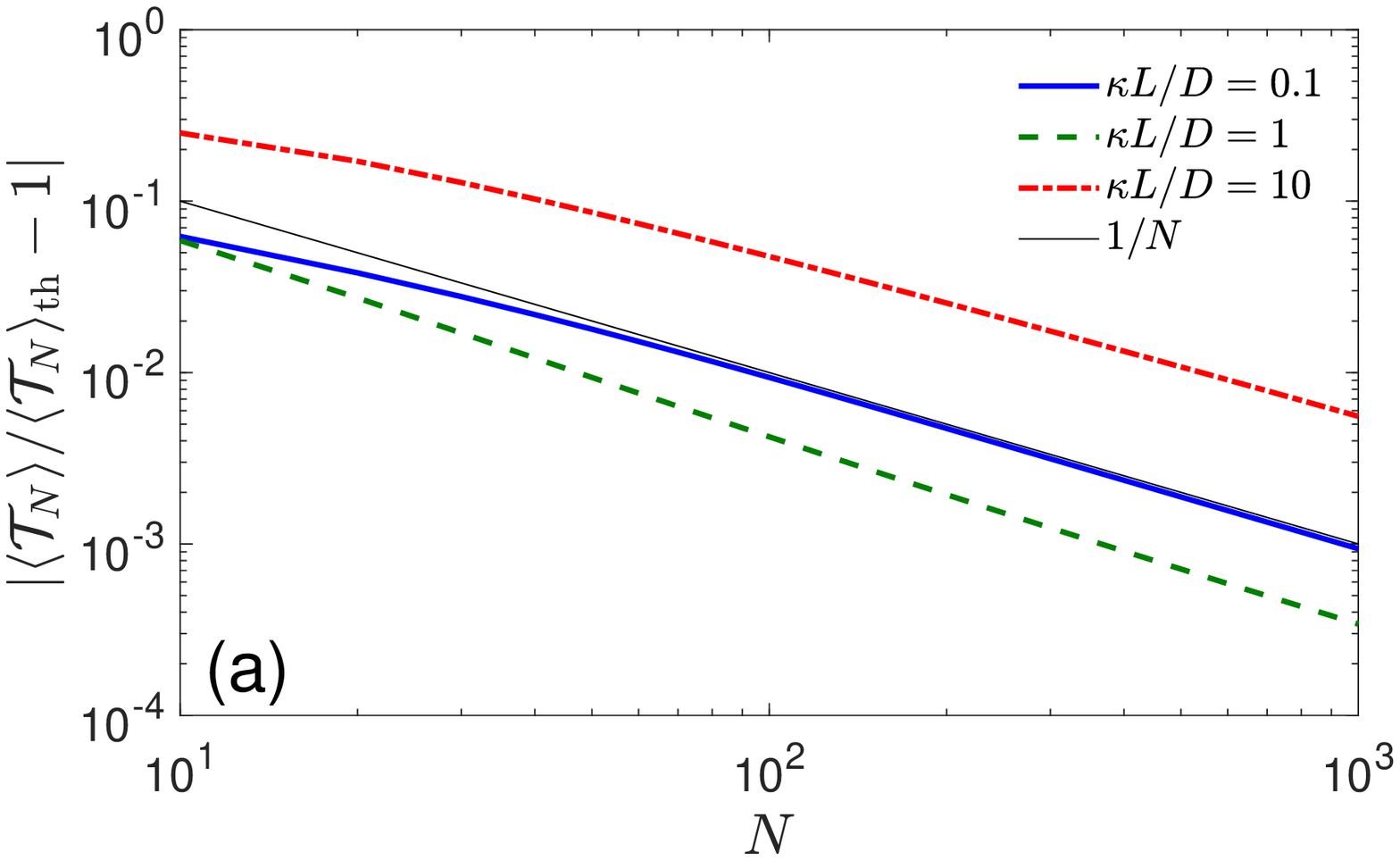} 
\includegraphics[width=88mm]{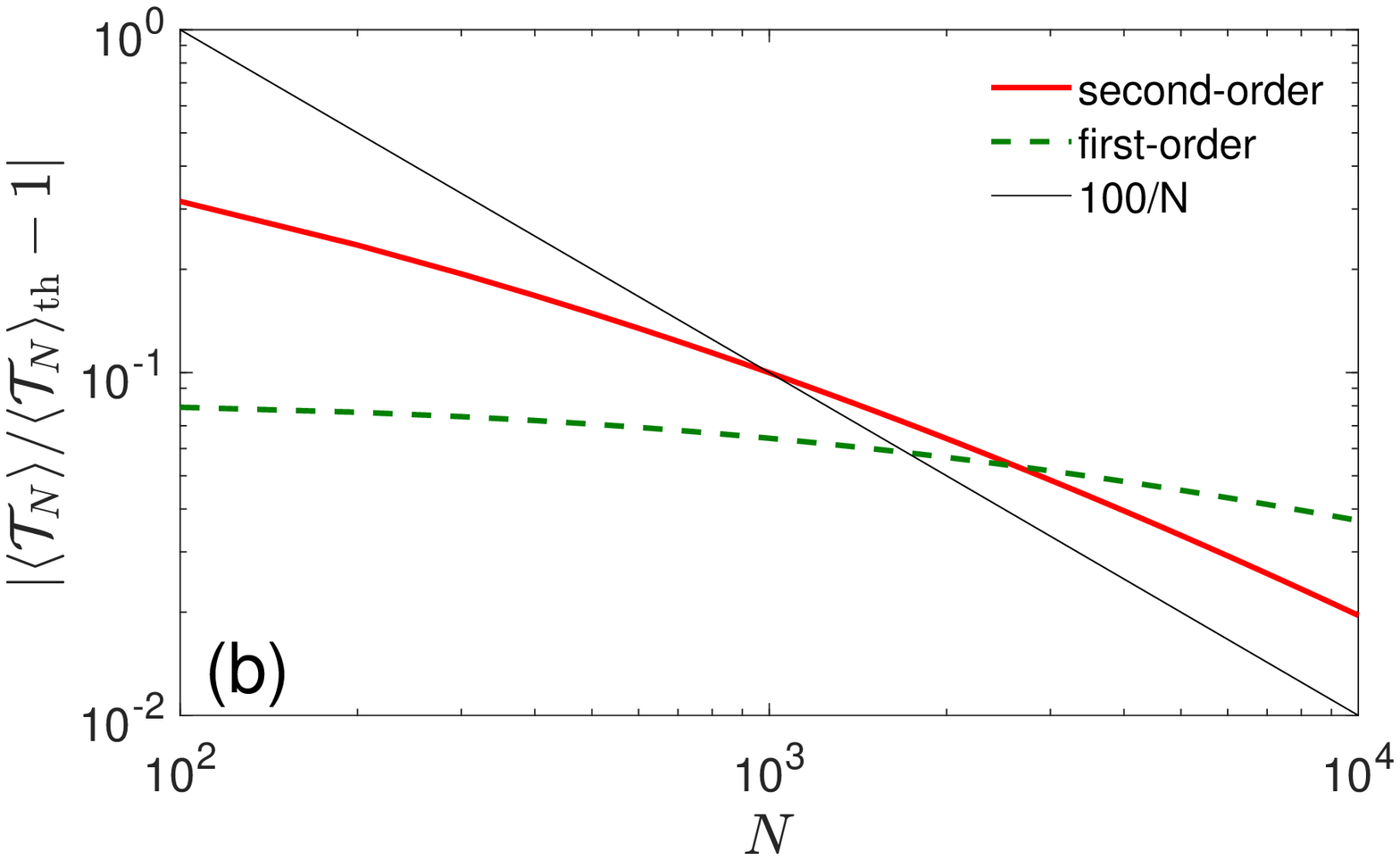} 
\caption{Absolute value of the relative error of the approximation
(\ref{eq:MfFPT_1d_approx2}) for the volume-averaged mean fFPT.  {\bf
(a)} The case of an interval $(0,L)$ with partially reactive
endpoint $L=1$ and reflecting endpoint $0$, with $D=1$ and the three
values of $\kappa L/D$ indicated in the plot. Lines represent the
ratio between $\overline{\langle\T_N\rangle}$ computed from numerical
evaluation of equation (\ref{eq:MfFPT_1d_volume}) and from the
approximation (\ref{eq:MfFPT_1d_approx2}). We checked numerically that
the relative error of the approximation decays as $O(N^{-1})$ (thin
black line).  {\bf (b)} The case of a spherical partially
reactive target of radius $R = 0.1 L$ surrounded by a reflecting
sphere of radius $L$, with $\kappa = 1$. The dashed line represents the
relative error obtained by using only the leading term in
(\ref{eq:MfFPT_1d_approx2}).}
\label{fig:rho1d_mean_kappa}
\end{figure}

Figure \ref{fig:rho1d_mean_kappa} illustrates the quality of this
approximation for several values of $\kappa$ for an interval of length
$L$ with a partially reactive endpoint (in which case $\Gamma =
\{ 0\}$ and thus $|\Omega|/|\Gamma| = L$).  The accuracy is lower
than in the former case of absorbing boundaries, because the error of
the approximation, $O(N^{-1})$, decreases slower than that of
Eq. (\ref{eq:MfFPT_1d_approx_new}).  In addition, we plot the relative
error in the case of a spherical partially reactive target of radius
$R = 0.1 L$ surrounded by a reflecting sphere of radius $L$.  In this
setting, the coefficient in front of the correction term,
$\sqrt{\kappa |\Omega|/(D|\Gamma|)} \approx 5.8$, is not small so that
this correction term turns out to deteriorate the quality of the
approximation at small $N$, as can be seen from the dashed line.  In
turn, when $N$ is large, the correction term improves the accuracy, as
expected.

Since the ratio $N/|\Omega|$ in (\ref{eq:MfFPT_1d_approx2}) can be
interpreted as a mean concentration $[A]$ of diffusing particles, our
result is consistent with standard chemical kinetics, in which the
reaction rate scales linearly with $[A]$.  Note, however, that we also
found the next-order term, which behaves as $\sqrt{[A]}$.
In contrast, expression (\ref{eq:MfFPT_1d_approx_new}) vanishes as
$1/N^2$ and hence, is proportional to a squared concentration of
particles. This is very different from the Smoluchowski rate which is
linear with the concentration, because it appears as a limiting form
when both the volume and the number of particles grow to infinity,
while their ratio is kept fixed and is sufficiently small (see
Appendix \ref{Smol}). It means that, in contrast to a standard
Collins-Kimball relation, in which both rate controlling factors enter
with the same power of concentration, for bounded domains in the limit
$N \to \infty$ the reaction inevitably enters into a regime of kinetic
control, while the controlling factor of diffusion becomes subdominant
and defines only some finite corrections. In a way, this situation is
similar to a transition to kinetic control upon reduction of the size
of the escape window, which was predicted recently for the so-called
narrow escape problem \cite{GrebenkovFPT,GrebenkovNEP}.

\subsection{Variance of the fFPT}
\label{sec:variance}

As the fFPT $\T_N$ includes two sources of randomness (from
Brownian trajectories and from random initial positions), there are at
least two ways of characterising the fluctuations of the $\T_N$. In the
first way, one may use the variance defined as
\begin{equation}
\label{eq:variance1}
\Var\{\T_N\}=\overline{\langle\T_N^2\rangle}-\bigl[\overline{\langle\T_N
\rangle}\bigr]^2,
\end{equation}
where the second term is just the square of the mean volume-averaged fFPT.
From a physical point of view, it is more convenient to consider the
volume-averaged conditional variance
\begin{equation}
\label{eq:variance2}
\overline{\var\{\T_N\}}=\overline{\bigl(\langle\T_N^2\rangle-\langle\T_N
\rangle^2\bigr)}.
\end{equation}
In other words, one first evaluates the variance of $\T_N$ for a fixed set
of initial positions $\x_1,\ldots,\x_N$, and then averages this conditional
variance over these positions distributed uniformly in $\Omega$.

To find both variances, we first evaluate the second moment of $\T_N$
with respect to the probability density $\rho_N(t|\x_1,\ldots,\x_N)$,
\begin{eqnarray}
\nonumber
\langle\T_N^2\rangle&=&\int\limits_0^\infty dt\, t^2 \, \rho_N(t|\x_1,\ldots,\x_N)\\  
&=&2\int\limits_0^\infty dt \, t \, S_N(t|\x_1,\ldots,\x_N),
\end{eqnarray}
and then average it over the starting points to find
\begin{equation}
\label{eq:TN2_1st}
\overline{\langle\T_N^2\rangle}=2\int\limits_0^\infty dt \, t \, \left[\overline{S(t)}
\right]^N.
\end{equation}
To get the variance in (\ref{eq:variance1}), it is sufficient to
subtract the squared mean $\overline{\langle \T_N\rangle}$ studied in
Sec. \ref{sec:mean}.  In turn, for the conditional variance in
(\ref{eq:variance2}), one needs to subtract the squared first moment
$\langle\T_N\rangle$, averaged over the starting points,
\begin{eqnarray} 
\nonumber
\overline{\langle\T_N\rangle^2}&=&\int\limits_0^\infty dt_1\int\limits_0^\infty dt_2  
\overline{S(t_1|\x_1,\ldots,\x_N) S(t_2|\x_1,\ldots,\x_N)}\\  
&=&\int\limits_0^\infty dt_1\int\limits_0^\infty dt_2\left[\overline{S(t_1)S(t_2)}
\right]^N.
\end{eqnarray}

In order to further simplify this expression we use the spectral expansions
(\ref{eq:S_x0}) and (\ref{eq:rho_x0}) along with the orthonormality of
Laplacian eigenfunctions to derive the identities
\begin{subequations}
\label{eq:identities}
\begin{eqnarray}
\nonumber
\overline{S(t_1)S(t_2)}&=&\sum\limits_nc_ne^{-D\lambda_n(t_1+t_2)}\\
\label{eq:S2t}
&=&\overline{S(t_1+t_2)},\\
\nonumber
\overline{\rho(t_1)S(t_2)}&=&\sum\limits_nD\lambda_nc_ne^{-D\lambda_n(t_1+t_2)}\\
\label{eq:rhoSt}
&=&\overline{\rho(t_1+t_2)},\\
\nonumber
\overline{\rho(t_1)\rho(t_2)}&=&\sum\limits_nD^2\lambda_n^2c_ne^{-D\lambda_n(
t_1+t_2)}\\
\label{eq:rho2t}
&=&\left(-\partial_t\overline{\rho(t)}\right)_{t=t_1+t_2}.
\end{eqnarray}
\end{subequations}
In particular, the first identity implies
\begin{equation}
\overline{\langle\T_1\rangle^2}=\sum\limits_n\frac{c_n}{(D\lambda_n)^2},
\end{equation}
which is twice smaller than $\overline{\langle \T_1^2\rangle}$, compare
equation (\ref{eq:T1k}).

Using the identity (\ref{eq:S2t}) we complete the above computation,
\begin{equation}
\overline{\langle\T_N\rangle^2}=\int\limits_0^\infty dt_1\int\limits_0^\infty
dt_2\left[\overline{S(t_1+t_2)}\right]^N,
\end{equation}
from which the volume-averaged conditional variance of the fFPT follows in the
form
\begin{equation}
\overline{\var\{\T_N\}}=2\int\limits_0^\infty dt \, t \, \left[\overline{S(t)}\right]^N
-\int\limits_0^\infty dt_1\int\limits_0^\infty dt_2\left[\overline{S(t_1+t_2)}
\right]^N.
\end{equation}

To proceed, we substitute again $\overline{S(t)}$ by its short-time approximation
(\ref{eq:Svol_short}). For a perfectly absorbing boundary ($\kappa=\infty$) we get
from equation (\ref{eq:TN2_1st}),
\begin{eqnarray}
\nonumber
\overline{\langle\T_N^2\rangle}&\simeq&2\int\limits_0^T dt \, t \, \bigl(1-B\sqrt{t}\bigr)^N 
=\frac{4}{B^4}\int\limits_{1-B\sqrt{T}}^1dx(1-x)^3x^N\\  \label{eq:TN2}
&\simeq&\frac{24}{B^4(N+1)(N+2)(N+3)(N+4)},
\end{eqnarray}
where $T$ is the cut-off time and $B=2|\Gamma|\sqrt{D}/[\sqrt{\pi}|\Omega|]$
(note that exponentially small corrections due to $(1-B\sqrt{T})^N$ are neglected
in Eq.~(\ref{eq:TN2})).  Substituting Eqs.~(\ref{eq:TN2}) and
(\ref{eq:MfFPT_1d_approx_new}) into Eq.~(\ref{eq:variance1}), we get
\begin{equation}
\Var\{\T_N\}\simeq\frac{20}{B^4}N^{-4}.
\end{equation}
Similarly, we get
\begin{eqnarray}
\nonumber
\overline{\langle\T_N\rangle^2}&\simeq&\int\limits_0^Tdt_1\int\limits_0^Tdt_2
\bigl(1-B\sqrt{t_1+t_2}\bigr)^N\\
\nonumber
&=&\frac{2\sqrt{2}}{B^4}\biggl(\int\limits_{1-B\sqrt{T}}^1dx(1-x)^3x^N\\
\nonumber
&&+\int\limits_{1-\sqrt{2}B\sqrt{T}}^{1-B\sqrt{T}}dx(1-x)(2B^2T-(1-x)^2)x^N\biggr)\\
&\simeq&\frac{12\sqrt{2}}{B^4(N+1)(N+2)(N+3)(N+4)}.
\end{eqnarray}
Combining these results, we get the volume-averaged conditional
variance of the fFPT in the leading in the limit $N \to \infty$ order:
\begin{equation}
\overline{\var\{\T_N\}}\simeq\frac{12(2-\sqrt{2})}{B^4}\, N^{-4}.
\end{equation}
Dividing this variance by the squared volume-averaged mean value of
the fFPT produces
\begin{equation}
\label{A}
\frac{\overline{\var\{\T_N\}}}{[\overline{\langle\T_N\rangle}]^2}\simeq \frac{3
(2-\sqrt{2})}{4}\approx0.44\qquad(N\gg 1).
\end{equation}
Similarly, one can also obtain the squared coefficient of variation of $\langle
\T_N\rangle_{\x_1,\ldots,\x_N}$ with respect to the random variables $\x_1,\ldots,
\x_N$,
\begin{equation}
\label{B}
\frac{\overline{\langle\T_N\rangle^2}}{[\overline{\langle T_N\rangle}]^2}-1
\simeq 3\sqrt{2}-1\approx3.24\qquad(N\gg 1).
\end{equation}

The following point is to be emphasised. In statistical analysis of
the first-passage phenomena in bounded domains
\cite{thiago,mejia,mejia2}, the coefficient of variation of the PDF is
a meaningful characteristic which probes its "effective"
broadness. Typically, in situations when this parameter is much less
than unity, one deals with a narrow distribution and the actual
behaviour is well-captured by its first moment---the mean
first-passage time, which sets a unique time scale. Only in this case
the first-passage times are "focused", i.e., concentrated around
the mean value. Conversely, when the coefficient of variation is of
order of unity or even exceeds it, the PDF, despite the fact that it
possesses moments of arbitrary order, exhibits in some aspects a
behaviour reminiscent of so-called broad distributions, like
heavy-tailed distributions which do not possess all moments. In this
case, fluctuations around the mean value are comparable to the mean
value itself and hence, it is most likely that the values of
first-passage times observed in two realisations of the process will
be disproportionately different. In the case at hand, we notice that
the coefficient of variation of the volume-averaged fFPT,
Eq. (\ref{A}), is of order of unity, which implies that here the
fluctuations of this quantity around its mean value are of order of
this value itself. More striking, the fluctuations of the fFPT
corresponding to randomly distributed initial positions exceed the
mean value of the fFPT, see Eq. (\ref{B}), which signifies that
randomness in the initial positions of the searchers has a very
pronounced effect on the spread of fFPTs. Therefore, we conclude that
for both volume-averaged and "bare" fFPTs no unique time scale exists
and their actual behaviour cannot be fully characterised by their mean
values.

\subsection{Long-time behaviour}
\label{sec:long-time}

Now we consider the volume-averaged probability density for $N$ particles. As
usual, the long-time limit presents the simplest setting for bounded domains
due to the exponential decay of the survival probability and the PDF, as well
as their volume averages. In particular one gets from equation (\ref{eq:S_vol_def})
\begin{equation}
\label{rho}
\overline{\rho_N(t)}\simeq ND\lambda_1c_1^N\exp(-ND\lambda_1t)\qquad(t\to\infty),
\end{equation}
that is, the characteristic decay time $1/(D\lambda_1)$ for a single particle
gets simply reduced by the factor $N$. This steeper exponential decay
shifts the PDF to smaller times.  The decay time is a natural timescale
of the diffusive exploration of the whole domain. For a single particle,
this timescale is usually close to the mean FPT.  Surprisingly enough,
this is not true in case when multiple particles are searching for perfect
targets starting from distinct random locations.  As evidenced by our result
(\ref{eq:MfFPT_1d_approx_new}), here the mean fFPT scales as $1/N^2$, i.e.,
it is much smaller than $1/(D\lambda_1 N)$.  Therefore, the speedup of the
reaction kinetics due to a deployment of $N$ searchers starting at distinct
random positions appears to be really striking, as compared to a much more
modest logarithmic increase in the efficiency predicted in the case when
all of them start from the same point.

It is also instructive to compare the exponential decay in (\ref{rho})
to the asymptotic Weibull density (\ref{eq:Weibull}). For partially reactive
targets, $k=1$ and the functional form of (\ref{eq:Weibull}) generally agrees
with result (\ref{rho}), except for the coefficients. In contrast, for the
case of perfectly reactive targets, $k=1/2$, and (\ref{eq:Weibull}) exhibits
a stretched-exponential decay with time. The difference between these two
asymptotic behaviours stems from the order of how the limits are taken: the
Weibull density (\ref{eq:Weibull}) was derived for a fixed $t$ as $N\to\infty$
\cite{Madrid20}, whereas expression (\ref{rho}) corresponds to the limit $t\to
\infty$ with fixed $N$.

\subsection{Fluctuations between individual realisations}
\label{sec:fluctuations}

When the starting point $\x_0$ is random, the survival probability $S(t|\x_0)$
and the PDF $\rho(t|\x_0)$ for each fixed $t$ can be considered as random
variables themselves. This circumstance has been emphasised in the insightful
paper \cite{Evans11} studying a survival of an immobile target in presence
of diffusive traps. Moments of the survival probability regarded as
a random variable, and the difference between the typical and mean behaviour
was analysed \cite{lubensky,renn,monthus} for the problem of survival of a
diffusive particle in the presence of immobilised traps. The volume-averaged
PDF, $\overline{\rho(t)}$, is simply the mean of $\rho (t|\x_0)$. In order
to characterise fluctuations it is instructive to compute the variance of
$\rho(t|\x_0)$.

\subsubsection*{Single particle} 

We characterise fluctuations by the squared coefficient of variation
\begin{equation}
\label{eq:gamma1}
\gamma_1(t) =\frac{\overline{\rho^2(t)} - [\overline{\rho(t)}]^2}{[\overline{\rho(t)}]^2} \,,
\end{equation}
where the second moment can be evaluated using equation
(\ref{eq:rho2t}).  At long times, we get
\begin{equation}
\gamma_1(t)\to\frac{1}{c_1}-1\qquad(t\to\infty),
\end{equation}
where $c_1$ is defined by Eq. (\ref{eq:cn}).  For instance, for an
interval $(0,L)$ with absorbing endpoints, $c_1=8/\pi^2$ and thus
$\gamma(\infty)=\pi^2/8-1\approx0.23$ is of order of unity.

For short times, equation (\ref{eq:rhovol_short}) implies
$\overline{\rho^2(t)}\simeq(B/2^{7/2})t^{-3/2}$ with
$B=2\sqrt{D}|\Gamma|/[\sqrt{\pi}|\Omega|]$, and thus
\begin{equation}
\label{eq:gamma1_t0}
\gamma_1(t)\simeq\frac{\sqrt{\pi}|\Omega|}{16|\Gamma|\sqrt{2D t}}+O(1)\quad(t\to 0).
\end{equation}
In other words, the coefficient of variation diverges in this limit,
meaning that the volume-averaged density is no longer representative.
This is confirmed by Fig. \ref{fig:rho1d_uniform_1}(a) see below.

\subsubsection*{Multiple particles}

This illustrative computation can be extended to the case of $N$
particles. We get
\begin{eqnarray}
\nonumber
\overline{\rho_N^2(t)}&=&N\overline{\rho^2(t)}\left(\overline{S^2(t)}\right)^{N-1}\\
&&+N(N-1)\left(\overline{\rho(t) S(t)}\right)^2\left(\overline{S^2(t)}\right)^{N-2}.
\end{eqnarray}
Using the identities (\ref{eq:identities}), we find that the squared
coefficient of variation obeys
\begin{eqnarray}
\nonumber
1+\gamma_N(t)&=&\left(\frac{\overline{S^2(t)}}{[\overline{S(t)}]^2}\right)^{N-2} 
\biggl\{\frac{1}{N}\frac{\overline{\rho^2(t)}}{[\overline{\rho(t)}]^2}\frac{
\overline{S^2(t)}}{[\overline{S(t)}]^2}\\
&&+(1-1/N)\biggl(\frac{\overline{\rho(t)S(t)}}{\overline{\rho(t)}\cdot\overline{
S(t)}}\biggr)^2 \biggr\}.
\label{eq:gammaN}
\end{eqnarray}
At long times, the ratios entering this expression approach $1/c_1$ so
that we get
\begin{equation}
\label{eq:gamma_inf}
\gamma_N(t)\simeq c_1^{-N}-1\qquad(t\to \infty).
\end{equation}
The H\"older inequality, applied to the first (positive) eigenfunction
$u_1$, implies that
\begin{equation}
\label{eq:c1_Holder}
c_1\leq1.
\end{equation}
As a consequence, $\gamma_N(t)$ exponentially {\it diverges} with $N$
for large $t$.  In section \ref{sec:c1} below we explain this
paradoxical blow-up of fluctuations.

At short times, we find
\begin{equation}
\label{eq:gammaN_t0}
\gamma_N(t)\simeq\frac{\sqrt{\pi}|\Omega|}{16|\Gamma|\sqrt{2Dt}}N^{-1}-(1+1/N)
\frac12\qquad (t\to 0)
\end{equation}
(note that one can also get correction terms $O(1)/N$ from the first
expression).  As the number of particles increases, there are two
effects.  On one hand, the first (divergent) term is progressively
attenuated.  In other words, for any fixed $t$, taking $N$ large
enough we diminish $\gamma_N(t)$ and hence, get narrower distribution
of the random variable $\rho = \rho_N(t|\x_1,\ldots,\x_N)$ (here we
forget that $\rho_N$ is itself the probability density and consider it
as a random variable due to the randomness of $\x_1,\ldots,\x_N$).  On
the other hand, the second term increases and rapidly reaches a
constant contribution $1/2$ to the squared coefficient of variation.
Even though the second term comes with the negative sign, it cannot
render the coefficient of variation negative: indeed, the short-time
asymptotic relation (\ref{eq:gammaN_t0}) is only valid for $t$ small
enough such that the first term in (\ref{eq:gammaN_t0}) provides the
dominant contribution.

\begin{figure}
\centering
\includegraphics[width=88mm]{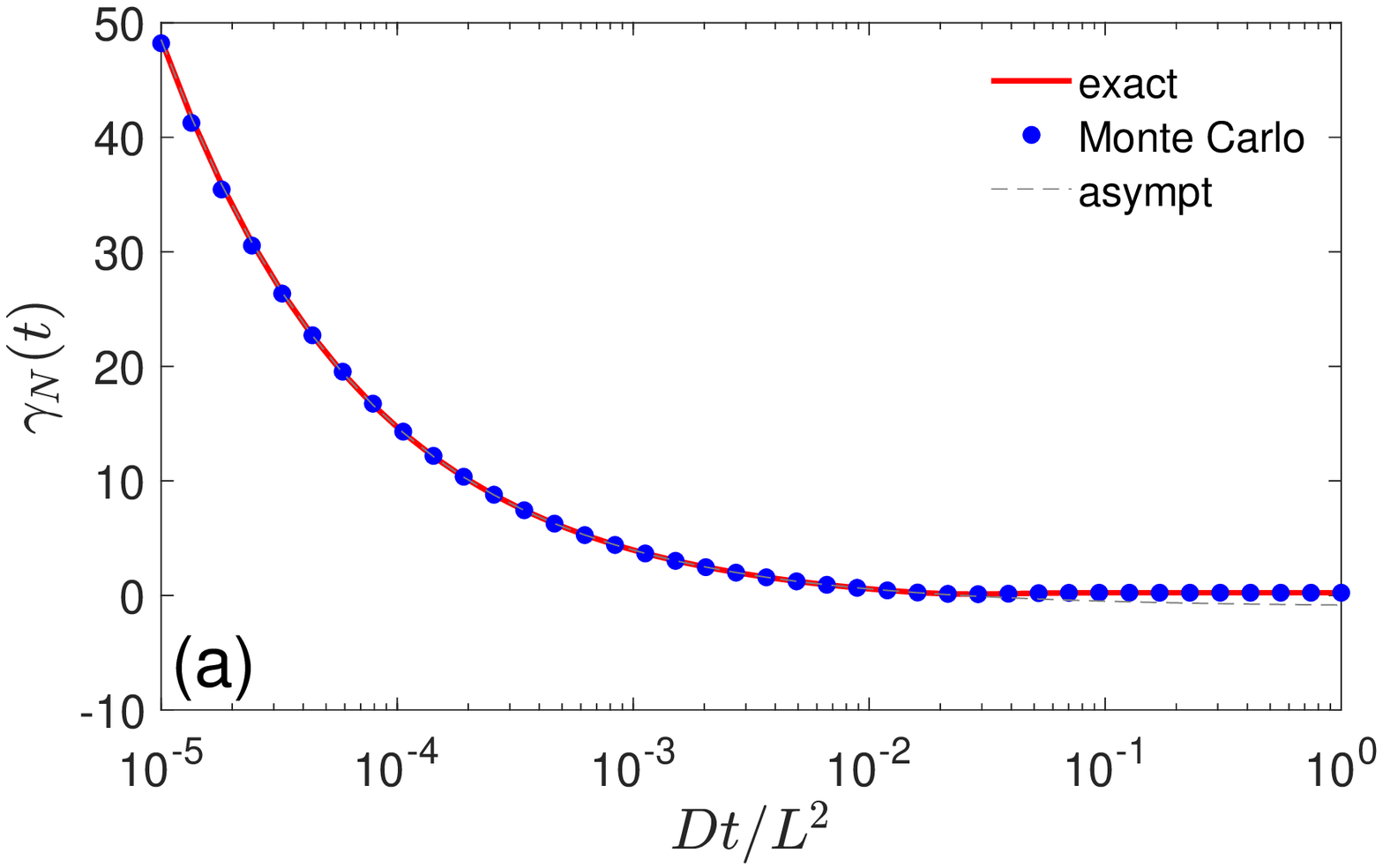} 
\includegraphics[width=88mm]{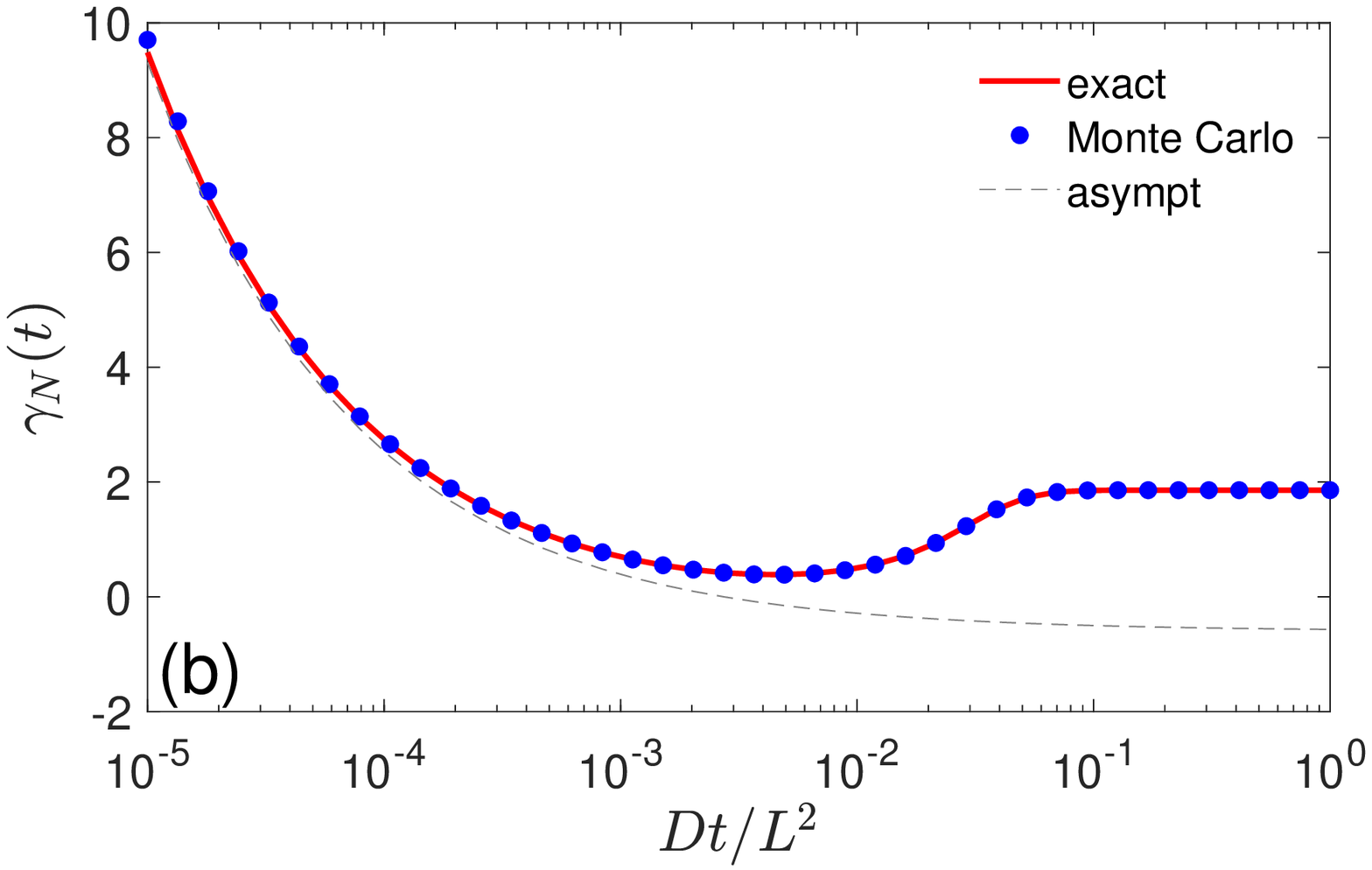} 
\includegraphics[width=88mm]{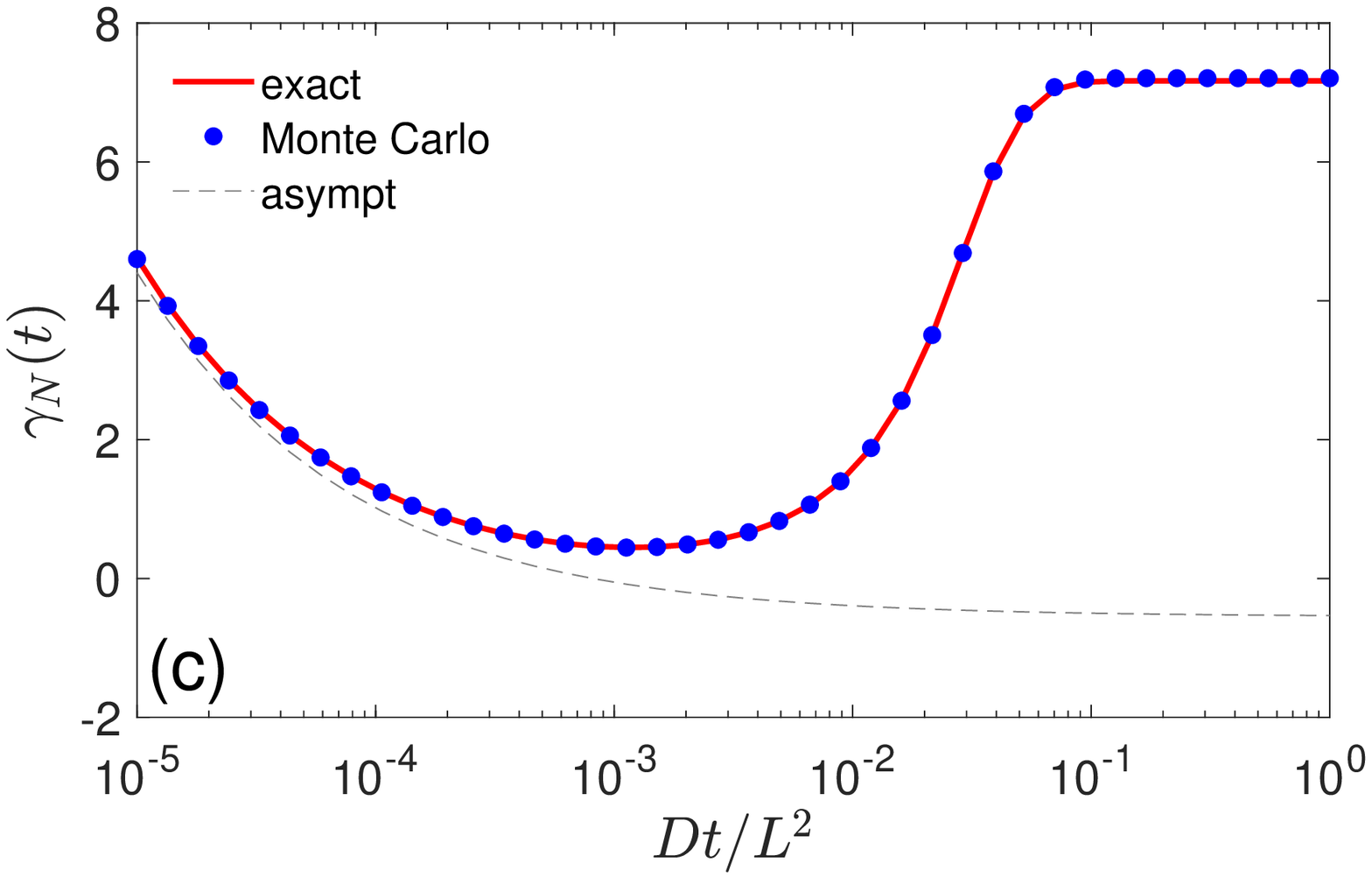} 
\caption{
Squared coefficient of variation $\gamma_N(t)$ for an interval $(0,L)$
with absorbing endpoints, with $L=1$ and $D=1$, and $N=1$ \textbf{(a)}, $N=5$
\textbf{(b)}, and $N=10$ \textbf{(c)}. The thick line represents the exact
solution (\ref{eq:gammaN}) (computed by truncating spectral expansions
at $n=1000$), the dashed line shows the short-time asymptotic
(\ref{eq:gammaN_t0}). Filled circles show the results of Monte Carlo
simulations with $M = 10^5$ trials.}
\label{fig:gamma1_1d}
\end{figure}

Figure \ref{fig:gamma1_1d} shows an excellent agreement between the
exact expression for $\gamma_N(t)$, its short-time asymptotic
behaviour and Monte Carlo simulations results for an interval
$(0,L)$. Note that we replaced the $O(1)$ term in equation
(\ref{eq:gamma1_t0}) by $-1$ because for an interval, the short-time
asymptotic formulas (\ref{eq:rhovol_short}) are exponentially
accurate, and the only correction $-1$ comes from the definition of
the coefficient of variation.

\begin{figure*}
\centering
\includegraphics[width=88mm]{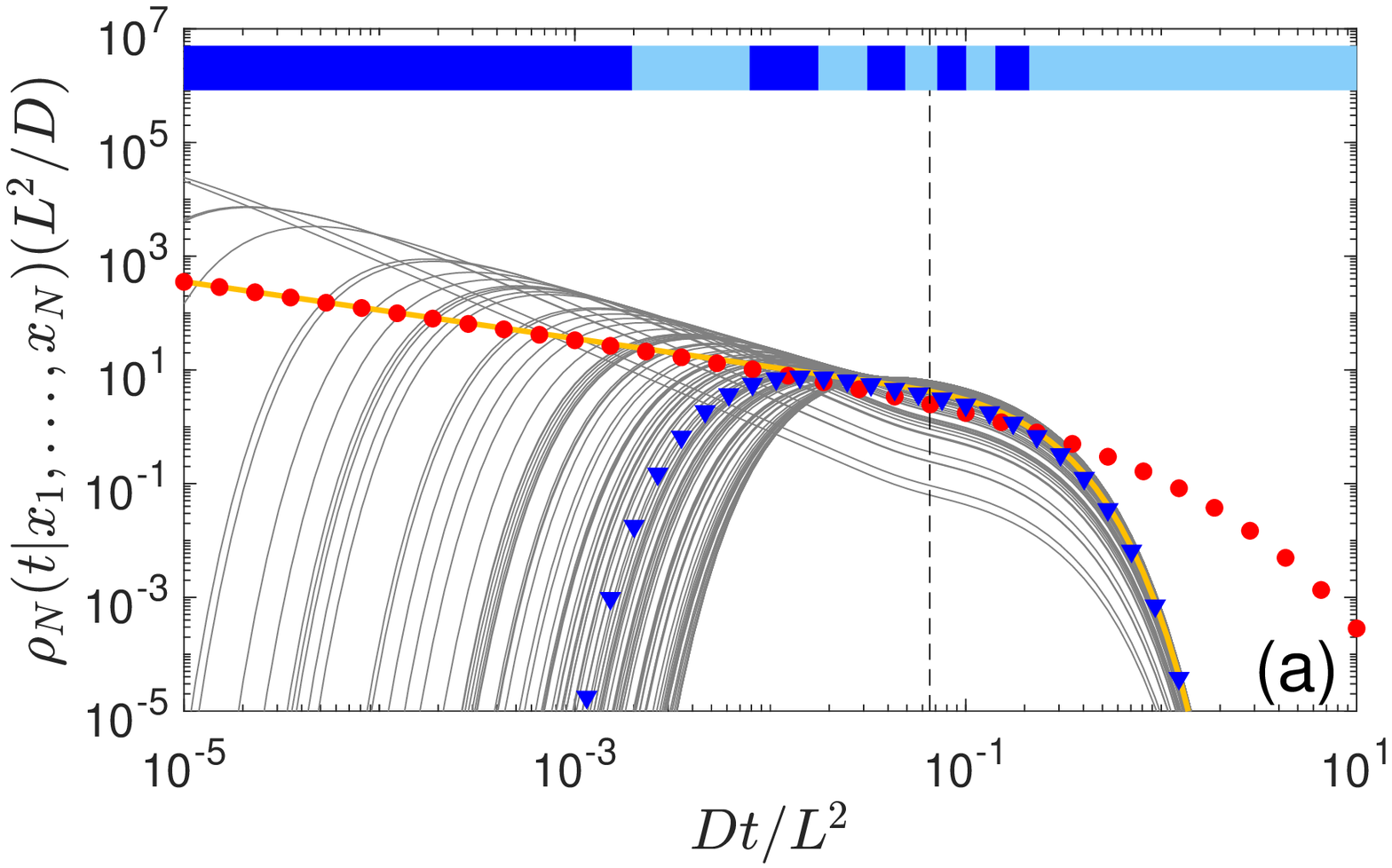} 
\includegraphics[width=88mm]{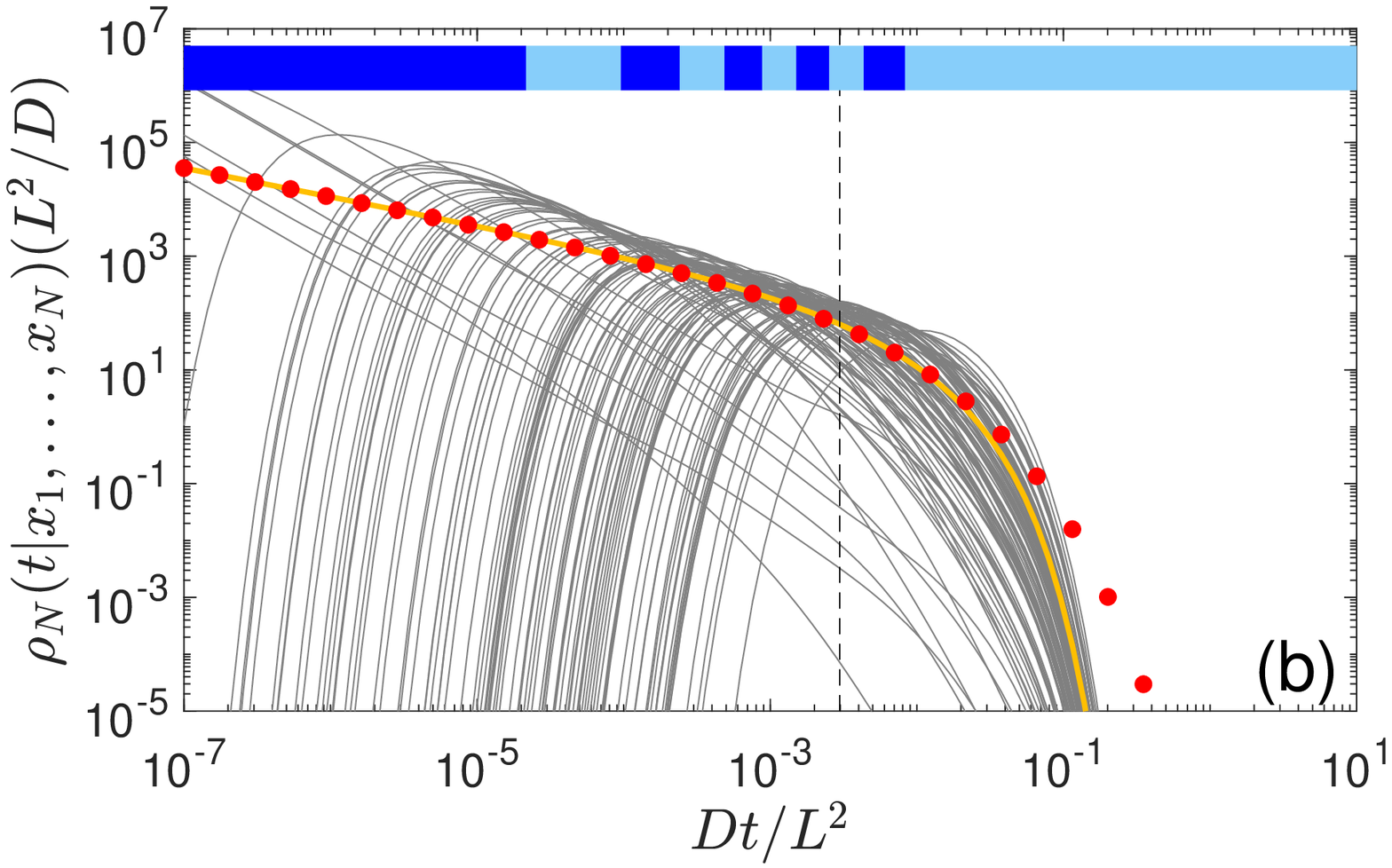} 
\includegraphics[width=88mm]{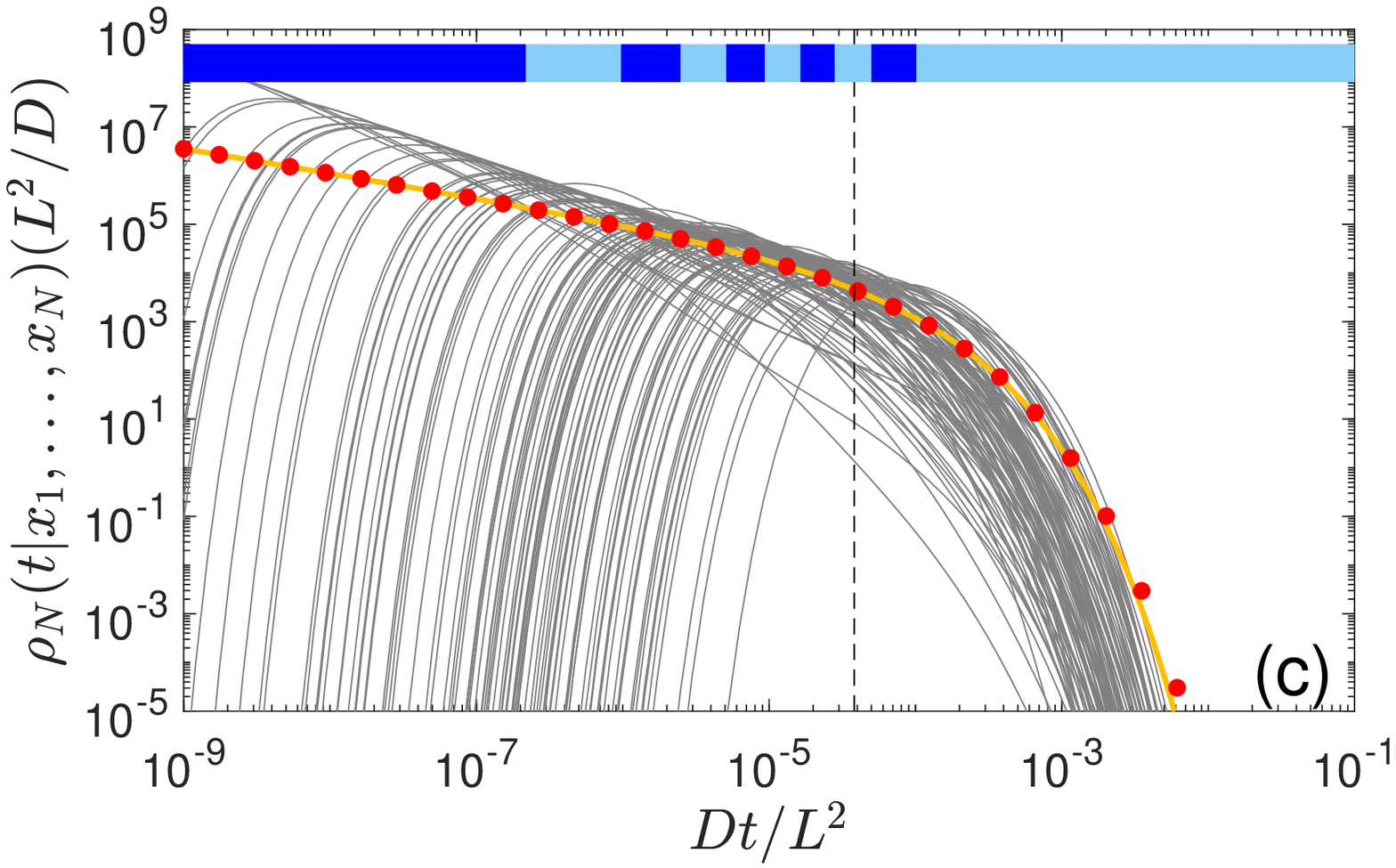} 
\includegraphics[width=88mm]{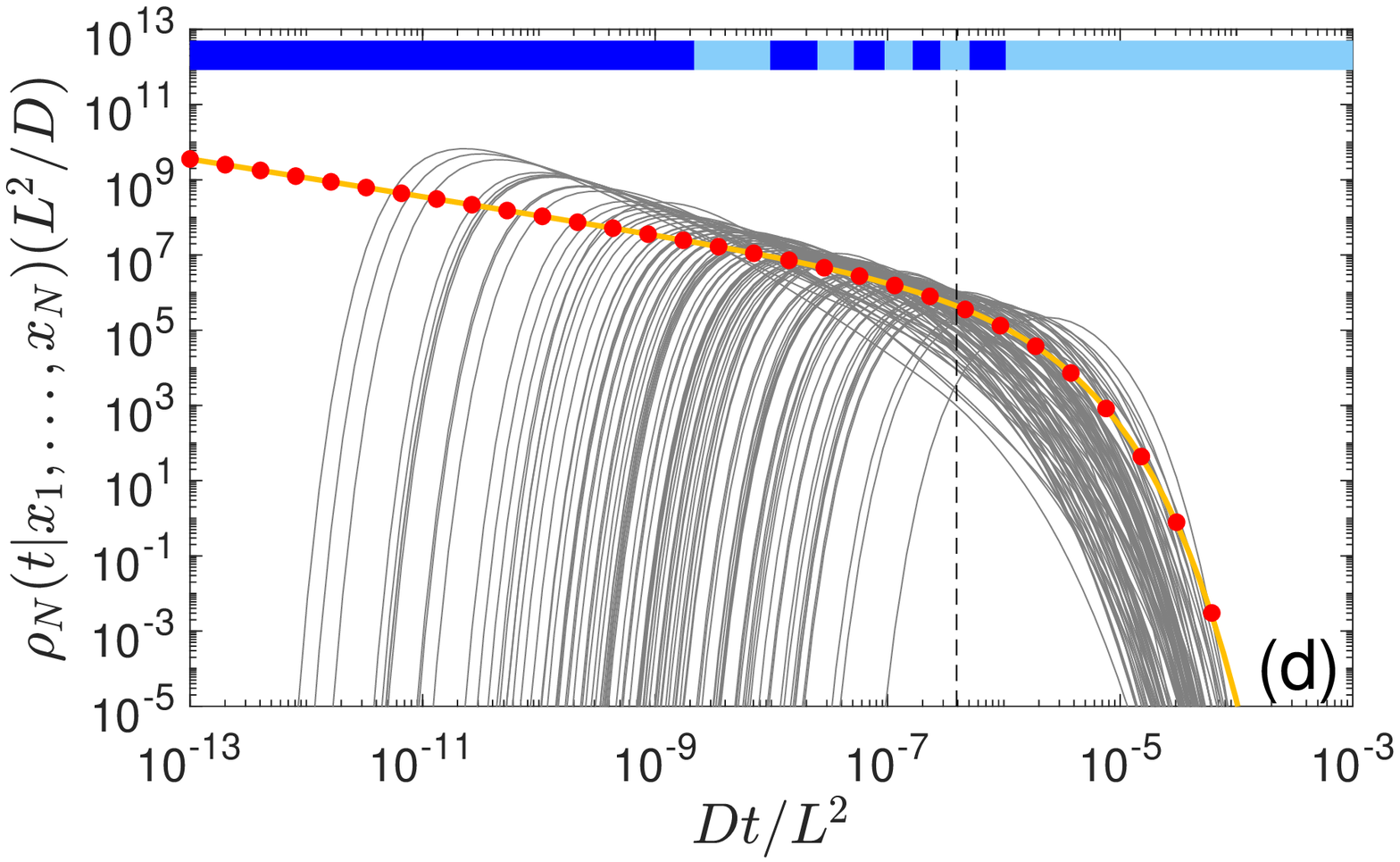} 
\caption{Probability density of the FPT for an interval $(0,L)$ with perfectly
absorbing endpoints ($\kappa = \infty$), $L=1$ and $D=1$.  The thin
grey lines represent for 100 realisations of the PDFs
$\rho_N(t|x_1,\ldots, x_N)$ with uniformly and independently chosen
starting point $x_1,\ldots,x_N$. The thick yellow line shows the
volume-averaged PDF $\overline{\rho_N(t)}$ from equation 
(\ref{eq:rho_N_vol}) with $\overline{\rho(t)}$ and $\overline{S(t)}$
given by (\ref{eq:rho1d_volume}) and (\ref{eq:S1d_volume}), whereas red
filled circles represent the asymptotic Weibull density
(\ref{eq:Weibull}).  In turn, filled blue triangles show
the typical PDF $\rho_{1,\rm typ}(t)$ from equation
(\ref{eq:rho_typ}), computed by numerical integration over the
starting point. {\bf (a)} $N=1$, {\bf (b)} $N=10$, {\bf (c)} $N=100$,
and {\bf (d)} $N=1000$. The vertical dashed line shows the
volume-averaged mean fFPT from equation (\ref{eq:MfFPT_1d_approx}).
Coloured bars at the top present ten $10\%$-quantiles based on
$\overline{S_N(t)}$.}
\label{fig:rho1d_uniform_1}
\end{figure*}

\begin{figure*}
\centering
\includegraphics[width=88mm]{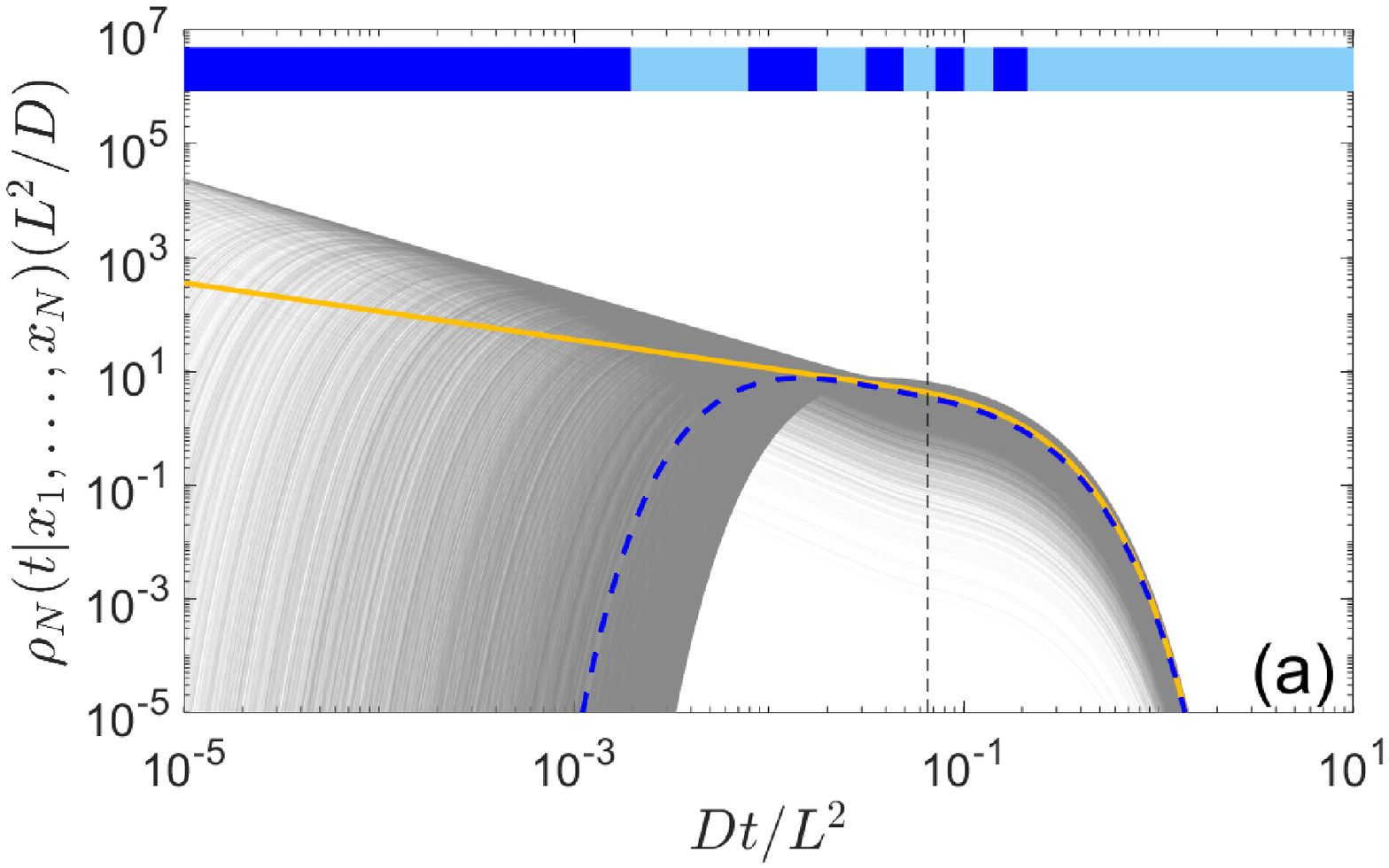} 
\includegraphics[width=88mm]{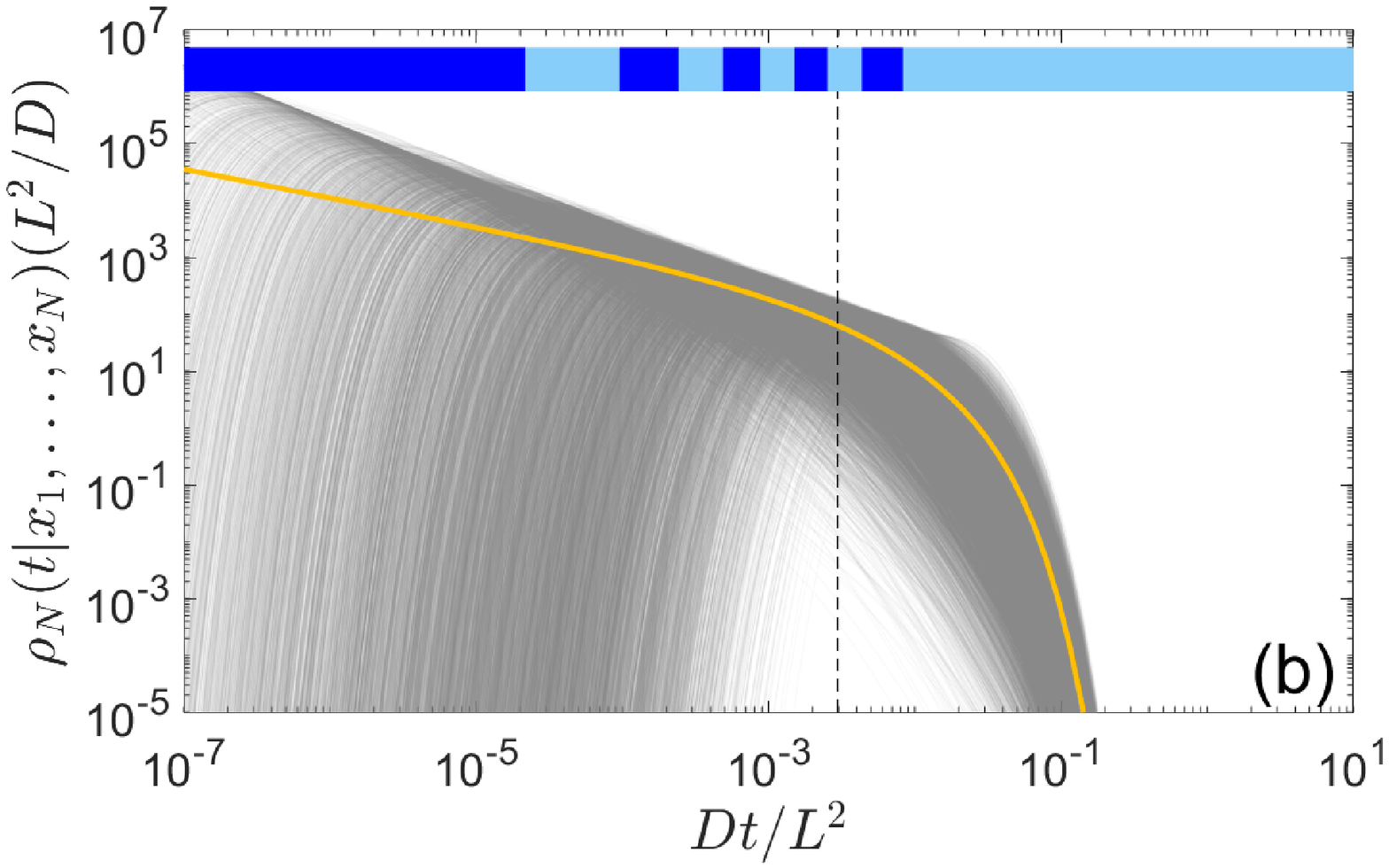} 
\includegraphics[width=88mm]{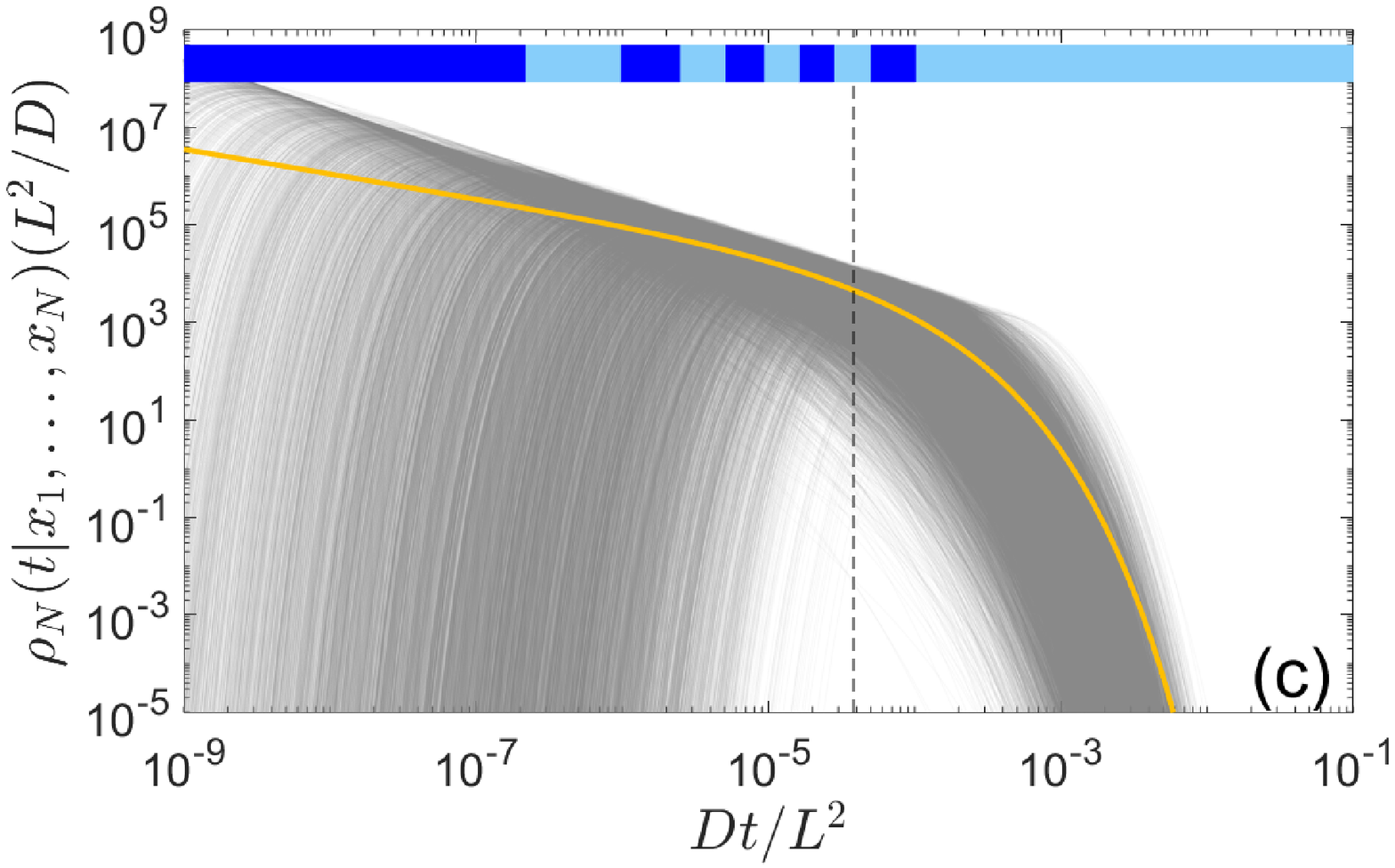} 
\includegraphics[width=88mm]{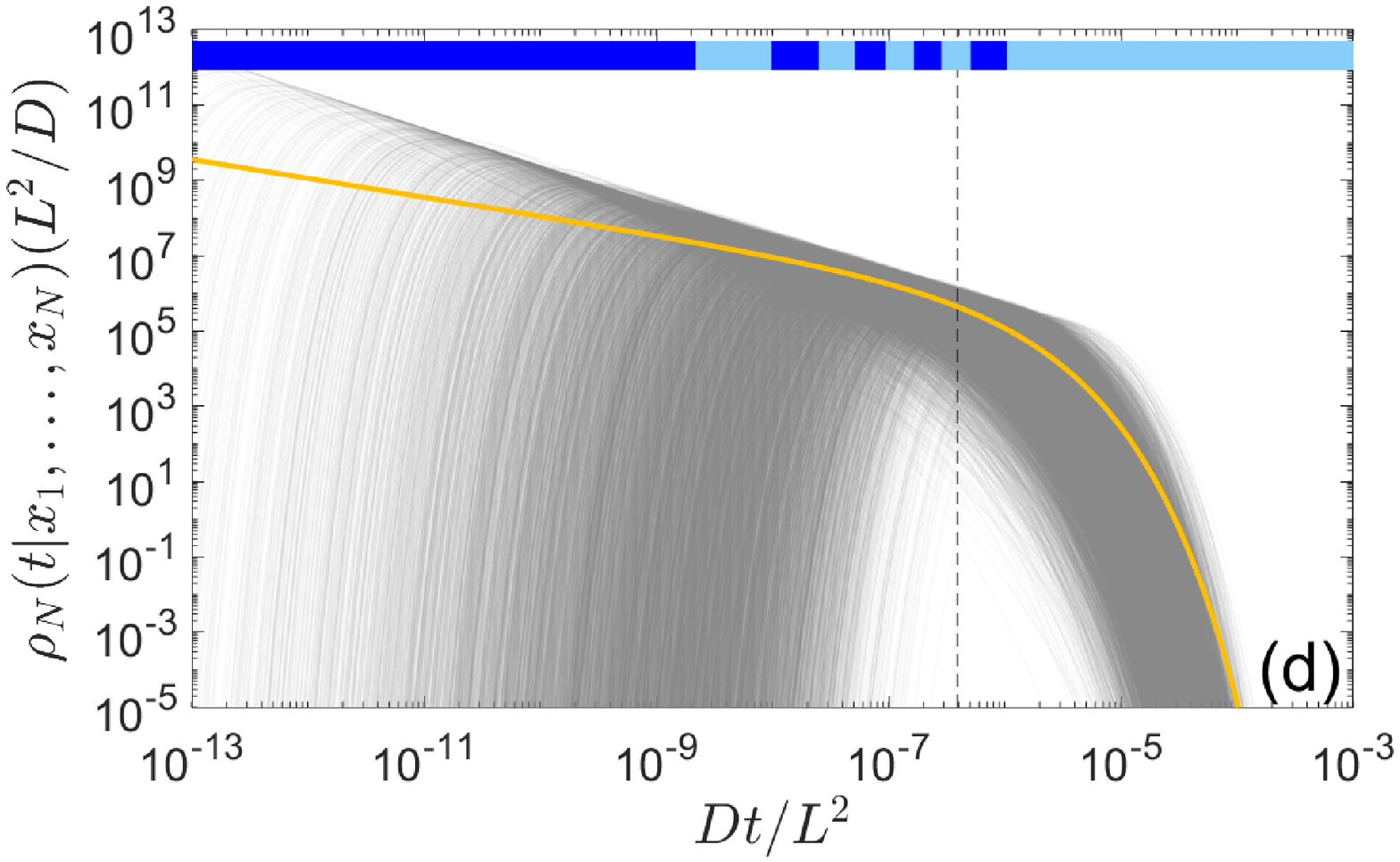} 
\caption{
The same as figure \ref{fig:rho1d_uniform_1} but with $10,000$
realisations of the PDFs $\rho_N(t|x_1,\ldots, x_N)$.}
\label{fig:rho1d_uniform_1bis}
\end{figure*}

Figure \ref{fig:rho1d_uniform_1} illustrates the PDFs
$\rho_N(t|x_1,\ldots,x_N)$ for 100 random combinations of the starting
points $x_1,\ldots,x_N$ chosen uniformly and independently on the
interval $(0,L)$.  Fluctuations of the PDFs $\rho_N(t|x_1,\ldots,x_N)$
around their volume-averaged mean $\overline{\rho_N(t)}$ are present
and significant for all $N$ ranging from $1$ to $1000$.  The broadness
of these fluctuations can be even better seen on figure
\ref{fig:rho1d_uniform_1bis} which illustrates $10,000$ random
realisations of the starting points.  This is a rather
counter-intuitive observation as a sort of convergence to the
volume-averaged PDF is expected as $N$ increases. Note that the
asymptotic Weibull density (\ref{eq:Weibull}) accurately describes the
volume-averaged mean $\overline{\rho_N(t)}$ already for $N=100$.

We emphasise that the PDF is rapidly shifted towards smaller times as $N$
increases. Intuitively, one could expect that this shift is controlled by
the long-time exponential decay and its timescale, $T_1/N$, see section
\ref{sec:long-time}. However, the appropriate time scale here is the
volume-averaged mean fFPT, $\overline{\langle\T_N\rangle}$, which decays
much faster, as $1/N^2$, see the vertical dashed lines. We stress again
that the decay time here is not related to the mean fastest FPT.

The general shape of the PDFs $\rho_N(t|\x_1,\ldots,\x_N)$ is significantly
different from the volume-average $\overline{\rho_N(t)}$ in figure
\ref{fig:rho1d_uniform_1}. Thus for a single realisation the starting point
is a finite distance away from the target, which effects the
exponential cutoff to very short times and the emergence of the peak
(the most probable FPT). At longer times we see the exponential
shoulder with time scale $T_1/N$. In figure \ref{fig:rho1d_uniform_1}
(a) individual realisations feature a slight bend at intermediate
times after the most probable FPT and the exponential shoulder,
however, this feature is getting lost for increasing $N$.  In this
one-dimensional setting the pronounced plateau in the FPT density
uncovered in \cite{Grebenkov18} is not present, a further separation
of the most probable time and the longest time scale $T_1/N$ would
build up for decreasing reactivity.  The volume-averaged PDF
$\overline{\rho_N(t)}$, in strong contrast, includes particles
starting arbitrarily closely to the target, and here we do not see the
initial exponential suppression. Instead $\overline{\rho_N(t)}$ decays
monotonically, and the only remaining relevant scale remaining is
$\overline{\langle\mathcal{T}_N\rangle}$. Nevertheless the FPT density
remains broad, as quantified by the 10\%-quantiles in figure
\ref{fig:rho1d_uniform_1} and our results for the coefficient of variation.

Note that the small $q$-quantiles can be approximated by using again the
short-time asymptotic formula (\ref{eq:Svol_short}).  In fact, writing
\begin{equation}
1-q=\overline{S_N(t)}=\left[\overline{S(t)}\right]^N\simeq(1-B\sqrt{t})^N,
\end{equation}
which is valid for small $t$ and thus small $q$, one gets easily
\begin{equation}
t_q\simeq\biggl(\frac{1-(1-q)^{1/N}}{B}\biggr)^2\simeq\frac{q^2}{B^2N^2}
\qquad(N\gg 1),
\end{equation}
where $B=2\sqrt{D}|\Gamma|/[\sqrt{\pi}|\Omega|]$. As can be seen
in figure \ref{fig:rho1d_uniform_1} the relative locations of the
quantiles hardly change for $N=10$, $100$, to $1000$ but are just
shifted to shorter times, as expected. This approximate relation is
indeed quite accurate. Here, the scaling is again as $1/N^2$, which is
a direct consequence of the leading term $O(\sqrt{t})$ in the
short-time expansion of the volume-averaged survival probability.

\subsection{Paradoxical divergence and its rationalisation} 
\label{sec:c1}

The constant $c_1$ from equation (\ref{eq:cn}) determines the behaviour of
the coefficient of variation in the long-time limit. This constant is defined
to be independent of the size of the domain, i.e., any dilation of the domain
does not change $c_1$. For instance, one has $c_1=8/\pi^2\approx0.81$ for an
interval with absorbing endpoints (of any length) and $c_1=6/\pi^2\approx0.61$
for an absorbing sphere (of any radius). According to inequality
(\ref{eq:c1_Holder}), this coefficient cannot exceed $1$, and it is actually
equal to $1$ only when the eigenfunction $u_1$ is constant, in other words,
for a reflecting boundary without any reaction. As a consequence, the
coefficient of variation $\gamma_N$ at long times exhibits an exponential
divergence with $N$. This observation appears paradoxical. We here resolve
this counter-intuitive behaviour.

First, we recall that assuming the limit $N\to\infty$ while keeping
the volume $|\Omega|$ fixed corresponds to a diverging concentration,
which is evidently unphysical. Moreover, in typical applications, the
diffusing particles search for a small target contained within a much
larger confining domain with reflecting boundary. In this setting, the
ground eigenfunction $u_1(\x)$ is close to a constant so that $c_1$ is
close to $1$. In other words, if the target is fixed but the confining
domain grows, $c_1$ tends to $1$, and the double limit $N\to\infty$
and $|\Omega| \to \infty$ with a fixed concentration $N/|\Omega|$
leads to a finite value of $\gamma_N$ and thus mends this paradoxical
divergence. To clarify these issues, it is therefore important to
investigate its behaviour in the small target limit. We first analyse
an exactly solvable setting and then briefly discuss the general case.

We consider the case of a spherical target of radius $R$ surrounded
by a larger reflecting sphere of radius $L$.  In Appendix
\ref{sec:targetS}, we provide the exact formula (\ref{eq:cn_sphere})
for the coefficients $c_n$, which depend on the solutions $\alpha_n$
of Eq. (\ref{eq:alphan_sphere}).  For small $R$ the smallest solution
$\alpha_0$ of this equation is expected to be small.  Denoting
$\epsilon = R/L \ll 1$, we consider separately the cases of infinite
and finite reactivity.

(i) For infinite reactivity ($\kappa=\infty$) one substitutes
$\alpha_0 \simeq a_1 \epsilon +a_2 \epsilon^2+O(\epsilon^3)$ into
equation (\ref{eq:alphan_sphere}) and expands it into a Taylor series
to determine the expansion coefficients $a_i$, from which we get
\begin{equation}  \label{eq:c1_sphere1}
c_1=1-\frac{108}{175}(R/L)^2+O(\epsilon^3).
\end{equation}
As a consequence, the coefficient of variation at long times can be
approximated as
\begin{equation}
\gamma_N(\infty)\simeq\exp\left(\pi\frac{144}{175} [A] R^2 L\right)-1,
\end{equation}
where $[A] = N/|\Omega|$ is the concentration of the diffusing
particles.

(ii) In case of a finite reactivity ($\kappa<\infty$) more terms are
needed. Substituting $\alpha_0 \simeq a_1 \epsilon + a_2 \epsilon^2 +
a_3 \epsilon^3 + a_4 \epsilon^4 + O(\epsilon^5)$ into
Eq. (\ref{eq:alphan_sphere}), we find
\begin{equation}
\label{eq:c1_sphere2}
c_1=1-\frac{108}{175}\frac{\kappa^2 R^4}{D^2 L^2}+O(\epsilon^5),
\end{equation}
and thus
\begin{equation}
\gamma_N(\infty)\simeq\exp\left(\pi\frac{144}{175}\frac{\kappa^2 R^4 L [A]}{D^2}
\right)-1.
\end{equation}
To ensure the validity of this asymptotic analysis, the first
correction term in equation (\ref{eq:c1_sphere2}) has to be small,
which is realised when the inequality $\kappa R^2 \ll D L$ holds. This
inequality is evidently valid in the low concentration limit when the
coefficient of variation vanishes, or when diffusion is very
fast. Moreover, it holds in case of a sufficiently low reactivity
$\kappa$; in this case, evidently, before a reaction eventually takes
place there are multiple encounters with a target interspersed with
bulk excursions such that the initial spatial heterogeneity due to a
random distribution particles gets speared away, effecting
$\gamma_N(\infty)$ to be small. In contrast, a larger domain favours a
higher degree of spatial heterogeneity and thus increases
$\gamma_N$. Note also that, in principle, one can relax this
restrictive relation between the parameters and use equations
(\ref{eq:cn_sphere}) and (\ref{eq:gamma_inf}) directly. This will
permit us to compute $c_1$ and hence, $\gamma_N(\infty)$ for any values
of the parameters.

In the above setting, the target was a small ball located in the
bulk and surrounded by a large reflecting sphere. If the target is
located on the outer sphere (e.g., a small circular hole), one has to
consider mixed Dirichlet-Neumann boundary condition on the sphere, for
which there is no exact solution for the survival probability.  In
this case, one can either resort to the approximate solution in
\cite{GrebenkovFPT}, or rely on the asymptotic analysis in the narrow
escape limit.  In the latter case, Ward and Keller showed for a
perfectly reactive target that
\begin{equation}  \label{eq:c1_Ward}
c_1 = 1 - E_1 \epsilon + o(\epsilon),
\end{equation}
where the coefficient $E_1$ was expressed as an integral of the
first-order correction to the ground eigenfunction \cite{Ward93}.
This coefficient can also be expressed in terms of the surface Neumann
Green function.  When the confining domain $\Omega$ is a ball, the
explicit form of this Green function was given in \cite{Cheviakov10}.
As a consequence, one can access the coefficient $E_1$ and thus the
asymptotic behaviour of $c_1$ in the narrow escape limit.  Moreover,
this computation is valid for multiple small targets located on the
sphere.  Importantly, the asymptotic relation (\ref{eq:c1_Ward})
remains valid even for nonspherical three-dimensional domains, even
though the computation of $E_1$ is much harder. We emphasise the
distinct scaling of $1-c_1$ with $\epsilon$: it is linear for a target
on the boundary, and quadratic for a target in the bulk, see
Eqs. (\ref{eq:c1_Ward}, \ref{eq:c1_sphere1}).  In two dimensions, the
approach of $c_1$ to $1$ is much slower:
\begin{equation}  \label{eq:c1_Ward_2d}
c_1 = 1 - \frac{E_1}{\ln (2/\epsilon)} + o( 1/\ln(2/\epsilon) ) ,
\end{equation}
where $2\epsilon L$ is the length of the target region, see
\cite{Ward93,Pillay10} for details.  An extension of these asymptotic
results to partially reactive targets located on the boundary presents
an interesting perspective.

\subsection{Typical PDF}
\label{sec:typical}

Following reference \cite{Evans11} we aim at determining the typical PDF
of the fastest FPT
\begin{equation}
\label{eq:rho_typ}
\rho_{N,\rm typ}(t)\equiv\frac{1}{\tau}\exp\biggl(\overline{\ln(\tau\rho_N(t|
\x_1,\ldots,\x_N))}\biggr),
\end{equation}
where $\tau$ is a timescale to render the expression in the logarithm
dimensionless. For this purpose, we need to compute
\begin{equation}\label{eq:auxil1}
\overline{\ln(\tau\rho_N(t|\x_1,\ldots,\x_N))}=U(t)+\sum\limits_{n=1}^N
\overline{\ln(S(t|\x_n))},
\end{equation}
where
\begin{equation}
U(t)\equiv\overline{\ln\bigl(\tau\rho(t|\x_1)/S(t|\x_1)+\ldots+\tau\rho(
t|\x_N)/S(t|\x_N)\bigr)}.
\end{equation}
As $\x_1,\ldots,\x_N$ are independent uniformly distributed starting points
the sum in equation (\ref{eq:auxil1}) contains $N$ identical terms.  In turn, the first
term $U(t)$ is more sophisticated as the logarithm couples different particles.
Using the following representation for the logarithm
\begin{equation}
\ln(p)=\int\limits_0^\infty\frac{dz}{z}\left(e^{-z}-e^{-pz}\right),
\end{equation}
we get
\begin{equation}
\label{eq:U_def}
U(t)=\int\limits_0^\infty\frac{dz}{z}\left(e^{-z}-[\eta(z;t)]^N\right),
\end{equation}
where
\begin{equation}
\label{eq:eta}
\eta(z;t)=\frac{1}{|\Omega|}\int\limits_{\Omega}d\x\exp\bigl(-z\tau\rho(t|
\x)/S(t|\x)\bigr).
\end{equation}

\subsubsection{Long-time behaviour}

The long-time behaviour is easy to determine. As $\rho(t|\x)=-\partial
S(t|\x)/\partial t$ the ratio $\rho(t|\x)/S(t|\x)$ is a positive
function that is monotonously increasing from $0$ at $t=0$ to
$D\lambda_1$ as $t\to\infty$. As a consequence equation
(\ref{eq:U_def}) implies
\begin{equation}
U(t)\to\ln(\tau ND\lambda_1)\qquad(t\to\infty). 
\end{equation}
Moreover, we have
\begin{equation}
\overline{\ln(S(t))}\simeq-Dt\lambda_1+b_1\qquad(t\to\infty),
\end{equation}
where
\begin{eqnarray}
\nonumber
b_1&=&\ln\biggl(|\Omega|^{-1/2}\int\limits_\Omega d\x \, u_1(\x)\biggr)\\
&&+\frac{1}{|\Omega|}\int\limits_{\Omega}d\x\ln\bigl(|\Omega|^{1/2}u_1(\x)\bigr),
\end{eqnarray}
and we recall that $u_1(\x)$ can be defined to be positive (factors $|\Omega|^{
\pm 1/2}$ were included to get dimensionless quantities in logarithms). In this
limit, one also has $\rho(t|\x_0)/S(t|\x_0)\simeq D\lambda_1$ so that
\begin{equation}
\rho_{N,\rm typ}(t)\simeq e^{b_1 N}ND\lambda_1e^{-Dt\lambda_1 N}\qquad(t\to\infty).
\end{equation}

In contrast, the analysis of the short-time behaviour of the
typical PDF is much more difficult and remains an open problem for
future research.

\section{Discussion}
\label{sec:discussion}

We studied the fFPT distribution for diffusion of $N$ Brownian
particles in a finite domain $\Omega$ with smooth boundary
$\partial\Omega$. The particles react with a surface region $\Gamma$
of perfect or partial reactivity contained in $\partial\Omega$. We
obtained the fFPT moments and probability density in the case of the
uniform initial condition. In the previously considered case of a
common, fixed initial position, the mean fFPT was shown to exhibit the
very slow, $1/\ln N$ scaling
\cite{Weiss83,Basnayake19,holcman,Lawley19,Lawley19b,Lawley19c}. In strong
contrast, when the $N$ particles are released random-uniformly in the
domain $\Omega$, the mean fFPT shows the much faster decay $1/N^2$ for
a perfectly reactive target and $1/N$ for a target with finite
reactivity.  In the former case, this scaling is also different from
that of the decay time, $T_1/N$. We provided a rationale of this
significantly altered behaviour in the dependence on the
specific initial condition by scaling arguments and supported it
by direct numerical computations.

While the scenario of a fixed initial condition bears relevance for
the example of the released sperm cells considered in \cite{Reynaud15,Basnayake19,
holcman}, spread-out initial conditions are relevant in other biological
systems. Thus, in biofilms autoinducer molecules are released on the cell
surfaces all over the colony \cite{biofilm,biofilm1}. In intracellular gene
regulation, so-called global regulatory proteins are distributed approximately
evenly throughout the cell volume \cite{kolesov}. Other regulatory proteins
may abound around their encoding gene \cite{kuhlman}, and genes that are
controlled by a specific transcription factor tend to locate next to the gene
encoding this transcription factor \cite{kepes,kolesov}. These examples
demonstrate the need for a theory that considers spread-out initial conditions
as considered here.

In many biological systems the number of searching entities such as
sperm cells or regulatory proteins are produced at the expense of
chemical and energy resources. At the same time a large number of
searchers guarantees a high degree of redundancy, which is of
particular importance for the case of sperm cells in singular, crucial
events such as reproduction. In contrast, for processes that are
constantly running off, such as gene regulatory processes, the
biochemical resources would quickly be drained if excessively many
molecules had to be produced. In this sense, and given the above
scenarios when the molecules abound in the vicinity of their target,
the scaling $1/N$ and $1/N^2$ provides a new perspective on why
relatively few molecules may already be sufficient to effect
comparatively speedy regulation.

While in our discussion we contrasted a fixed initial position with
the scenario of random-uniform initial positions, figure
\ref{fig:rho1d_uniform_1} indicates that the value of the mean
fFPT provides the correct scale even for individual random
realisations of starting points. In this sense, one may roughly
distinguish two classes of initial conditions: (i) either the
particles are allowed to start near their target (distributed in the
entire domain or concentrated in a subdomain around the target), (ii)
or the initial particle position is characterised by a minimal
distance to the target.  The former class corresponds to our analysis
here (see also \cite{Madrid20}), the latter class is described
by the results in previous works
\cite{Basnayake19,holcman,Lawley19,Lawley19b,Lawley19c}.

In summary, we believe that our results for the statistic of the fFPT
for spread initial particle positions provide a fresh perspective to the
field of many-particle search for an immobile target in a finite domain and its
applications. It should be of interest to further develop this approach both
from a mathematical point of view and with regard to concrete applications.

\begin{acknowledgments}
D.S.G. thanks M.J. Ward for fruitful discussions and acknowledges a
partial financial support from the Alexander von Humboldt Foundation
through a Bessel Research Award. R.M. acknowledges funding from the
German Science Foundation (DFG, grant ME 1535/7-1) as well as support
from the Foundation for Polish Science (Fundacja na rzecz Nauki
Polskiej, FNP) within an Alexander von Humboldt Polish Honorary
Research Scholarship.  G.O. is grateful to B. Meerson for helpful
discussions. The authors also thank the unknown referee for bringing
our attention to Ref.~\cite{Madrid20} that was published after the
submission of our manuscript and provided complementary views on
the fastest first-passage time.
\end{acknowledgments}

\appendix
\section{Smoluchowski limit}
\label{Smol}

In this Appendix we recall the behaviour in the conventional
thermodynamic limit when both $N$ and $|\Omega|$ tend to infinity with
the concentration $[A] = N/|\Omega|$ being fixed.  In this limit, one
gets
\begin{eqnarray*}
&& \overline{S_N} = \bigl[\overline{S(t)}\bigr]^N \simeq \biggl(1 - \frac{1}{|\Omega|} \int\limits_{\Omega} d\x_0 (1 - S(t|\x_0)) \biggr)^N \\
&& \simeq \exp\biggl(-[A] \int\limits_{\Omega} d\x_0 (1 - S(t|\x_0)) \biggr) = \exp(-[A] F(t)),
\end{eqnarray*}
where
\begin{equation}
F(t) = \int\limits_{\Omega} d\x_0 \bigl(1 - S(t|\x_0)\bigr) .
\end{equation}
For the survival probability of a spherical target of radius $R$ one
gets (see, e.g., \cite{Grebenkov10a})
\begin{eqnarray} 
\label{CK}
F(t) &=& 4\pi R \biggl[\frac{Dt}{1 + D/(\kappa R)} + \frac{2\sqrt{Dt} \, R}{\sqrt{\pi} (1 + D/(\kappa R))^2}  \\  \nonumber
&-& \frac{R D/\kappa}{(1+D/(\kappa R))^3} 
\biggl(1 - \erfcx\bigl(\sqrt{Dt}(1/R + \kappa/D)\bigr)\biggr)\biggr].
\end{eqnarray}
In the limit $t \to \infty$ the dominant behaviour of $F(t)$ is
provided by the first term, which yields the Collins-Kimball relation
\begin{equation}
\frac{t}{F(t)} = \frac{1}{4 \pi D R} + \frac{1}{K} \,,
\end{equation}
where $K = 4 \pi R^2 \kappa$ is the intrinsic reaction constant. 

In the limit of an infinitely large intrinsic reactivity ($\kappa\to
\infty$), one retrieves from Eq. (\ref{CK}) the classical
Smoluchowski result
\begin{equation}
F(t) = 4\pi R Dt \biggl(1 + \frac{2 R}{\sqrt{\pi Dt}}\biggr).
\end{equation}

\section{Improved formula for the mean fFPT}
\label{sec:mfFPT}

Our computation of the volume-averaged mean fFPT relied on the leading
term of the short-time expansion of the survival probability.  We can
improve this computation by accounting for the next-order correction.

\subsection{Perfect reactivity}

For the perfect reactivity, one has \cite{vandenBerg89,vandenBerg94}
\begin{equation}
S(t) \simeq 1 - B \sqrt{t} + C t + O(t^{3/2}) \qquad (t\to 0),
\end{equation}
where $B = (2/\sqrt{\pi}) \sqrt{D} \frac{|\Gamma|}{|\Omega|}$ and
\begin{equation}  \label{eq:R_def}
C = \frac{D|\Gamma|}{|\Omega| \RH} \,,
\qquad \frac{1}{\RH} = \frac{1}{|\Gamma|} \int\limits_{\Gamma} d\s \, H(\s),
\end{equation}
where $H(\s)$ is the mean curvature of the boundary at the point $\s$.
Substituting this expression into the first
integral in Eq. (\ref{eq:MfFPT_auxil}), we get
\begin{align*}
\overline{\langle \T_N\rangle} & \simeq \int\limits_0^T dt\, \bigl(1 - B \sqrt{t} + C t\bigr)^N 
= \int\limits_{z_0}^{1} dz \, f(z) \, z^N ,
\end{align*}
where we changed the integration variable as $z = 1-B\sqrt{t}+Ct$,
with $z_0 = 1 - B\sqrt{T} + CT$ and
\begin{equation*}
f(z) = \frac{B - \sqrt{B^2 - 4B_1(1-z)}}{C \sqrt{C^2 - 4B_1(1-z)}} \,.
\end{equation*}
For large $N$, the main contribution comes from the vicinity of $z =
1$ so that one can expand the function $f(z)$ into a Taylor series
around this point
\begin{equation*}
f(z) = f'(1) (z-1) + \frac12 f''(1) (z-1)^2 + O((z-1)^3) ,
\end{equation*}
and we used that $f(1) = 0$.  Integrating term by term, one gets
\begin{equation*}
\overline{\langle \T_N\rangle} \simeq \frac{-2f'(1)}{(N+1)(N+2)} + \frac{2f''(1)}{(N+1)(N+2)(N+3)} + \ldots,
\end{equation*}
with $f'(1) = -1/B^2$ and $f''(1) = 6C/B^4$.  As a consequence, we get
\begin{equation} \label{eq:Tmean_2nd}
\overline{\langle \T_N\rangle} \simeq \frac{2/B^2}{N^2} + \frac{12C/B^4 - 6/B^2}{N^3} + O(N^{-4}) ,
\end{equation}
which takes the form of Eq. (\ref{eq:MfFPT_1d_approx_new}).

\subsection{Partial reactivity}

For partial reactivity, one has \cite{Desjardins94}
\begin{equation}
S(t) \simeq 1 - B t + C t^{3/2} + O(t^2) \qquad (t\to 0),
\end{equation}
where 
\begin{equation}
B = \frac{\kappa |\Gamma|}{|\Omega|} \,, \qquad
C = \frac{4(\kappa/D)^2 |\Gamma|}{3\sqrt{\pi} |\Omega|} D^{3/2}  \,.
\end{equation}
Substituting this expression into the first integral in
Eq. (\ref{eq:MfFPT_auxil}), we get
\begin{align*}
\overline{\langle \T_N\rangle} & \simeq \int\limits_0^T dt\, \bigl(1 - B t + C t^{3/2}\bigr)^N 
= \int\limits_{z_0}^{1} dz \, f(z) \, z^N ,
\end{align*}
where we changed the integration variable as $z = 1-Bt+Ct^{3/2}$, with
$z_0 = 1 - B T + CT^{3/2}$, $f(z) = 1/(B - 3C \sqrt{t(z)}/2)$, and
$t(z)$ is the inverse of the above function $z(t)$.  Even though
$f(z)$ can be written explicitly in terms of the root of the cubic
polynomial, it is not needed as we only use its expansion around $z =
1$, which reads
\begin{equation*}
f(z) = \frac{1}{B} + \frac{3C}{2B^{5/2}} (1-z)^{1/2} + O(1-z) .
\end{equation*}
Integrating term by term, one gets
\begin{equation}
\overline{\langle \T_N\rangle} \simeq \frac{1}{B} N^{-1} + \frac{3\sqrt{\pi} C}{4B^{5/2}} N^{-3/2} + O(N^{-2}),
\end{equation}
which takes the form of Eq. (\ref{eq:MfFPT_1d_approx2}).

\section{Explicit solutions}

In this Appendix, we summarise the well-known explicit solutions for
two simple settings that we used for illustrations: an interval and a
sphere.

\subsection{Interval with absorbing endpoints}
\label{sec:1D_model}

For the interval $(0,L)$ with absorbing endpoints at $0$ and $L$, the
survival probability admits two equivalent explicit expansions
\begin{subequations}
\begin{eqnarray}
\label{eq:St_1d}
&& S(t|x_0) = 1-\erfc\biggl(\frac{x_0}{\sqrt{4Dt}}\biggr)\\  \nonumber  
&& \quad +\sum\limits_{k=1}^\infty(-1)^k\biggl[\erfc\biggl(\frac{kL-x_0}{\sqrt{4Dt}}
\biggr)-\erfc\biggl(\frac{kL+x_0}{\sqrt{4Dt}}\biggr)\biggr]\\
&& \quad =2\sum\limits_{k=1}^\infty\sin(\pi kx_0/L)\frac{1-(-1)^k}{\pi k} e^{-\pi^2k^2Dt/L^2}.
\end{eqnarray}
\end{subequations}
The first relation is numerically more convenient at small $t$, whereas the
second at large $t$. We also have
\begin{subequations}
\begin{eqnarray}
&&\rho(t|x_0)=\frac{1}{\sqrt{4\pi Dt^3}}\biggl[x_0e^{-x_0^2/(4Dt)} \\  \nonumber
&& \quad +\sum\limits_{k=1}^\infty(-1)^{k+1}
\biggl((kL-x_0)e^{-(kL-x_0)^2/(4Dt)}\\  \nonumber
&& \quad -(kL+x_0)e^{-(kL+x_0)^2/(4Dt)}\biggr)\biggr]\\
&& \quad =\frac{2D}{L^2}\sum\limits_{k=1}^\infty\sin(\pi kx_0/L)(1-(-1)^k) \\  \nonumber
&& \quad \times (\pi k) e^{-\pi^2k^2Dt/L^2}.
\end{eqnarray}
\end{subequations}
Note also that the Laplace-transformed probability density reads 
\begin{equation}
\tilde{\rho}(p|x_0)=\frac{\sinh(x_0\sqrt{p/D})+\sinh((L-x_0)\sqrt{p/D})}{
\sinh(L\sqrt{p/D})},
\end{equation}
from which one can easily compute positive-order moments.

The volume-averaged probability density is then
\begin{subequations}
\begin{eqnarray}
\label{eq:rho1d_volume}
\overline{\rho(t)}&=&\frac{2\sqrt{D}}{L\sqrt{\pi t}}\biggl(1+2\sum\limits_{k=1}^
\infty(-1)^ke^{-(kL)^2/(4Dt)}\biggr)\\
&=&\frac{8D}{L^2}\sum\limits_{k=1}^\infty e^{-\pi^2(2k-1)^2Dt/L^2},
\end{eqnarray}
\end{subequations}
and its Laplace transform of this density reads
\begin{equation}
\L\{ \overline{\rho(t)}\}=\frac{2}{z}\biggl(\frac{2\cosh z}{\sinh z}-\frac{\cosh(
z/2)}{\sinh(z/2)}\biggr),
\end{equation}
where $z=L\sqrt{p/D}$. One also has
\begin{subequations}
\begin{eqnarray}
\nonumber
\overline{S(t)}&=&1-\frac{4\sqrt{Dt}}{L\sqrt{\pi}}\biggl(1+2\sum\limits_{k=1}^
\infty(-1)^k\biggl(e^{-(kL)^2/(4Dt)}\\
\label{eq:S1d_volume}
&&-\frac{\sqrt{\pi}kL}{\sqrt{4Dt}}\erfc(kL/\sqrt{4Dt})\biggr)\biggr)\\
&=&\frac{8}{\pi^2}\sum\limits_{k=1}^\infty\frac{e^{-\pi^2(2k-1)^2Dt/L^2}}{(2k-1)^2}.
\end{eqnarray}
\end{subequations}
For an interval $(0,L)$, one has $\lambda_n=\pi^2n^2/L^2$, $u_n(x)=\sqrt{2/L}\sin(
\pi nx/L)$, and $c_n=2(1-(-1)^n)^2/(\pi^2 n^2)$.

We also easily get
\begin{equation}
\overline{\delta^2(\x_0)}=\frac{L^2}{12}
\end{equation}
and
\begin{equation}
b_1=\frac{1}{L}\int\limits_0^L dx\ln\biggl(\sqrt{2}\sin(\pi x/L)\biggr)=-\frac12
\ln2.
\end{equation}

\subsection{Interval with a partially reactive endpoint}

We also consider the interval $(0,L)$ with partially reactive endpoint at $L$
and reflecting endpoint at $0$, for which the Laplacian eigenvalues and
eigenfuctions are (see, e.g., \cite{Grebenkov17e})
\begin{equation}
\lambda_n=\alpha_n^2/L^2,\qquad u_n(x)=\beta_n\cos(\alpha_nx/L),
\end{equation}
with
\begin{equation}
\beta_n=\biggl(\frac{h^2+\alpha_n^2}{h^2+h+\alpha_n^2}\biggr)^{1/2} ,
\end{equation}
and $\alpha_n$ are the positive zeros of the equation $\alpha_n\sin\alpha_n=h
\cos\alpha_n$, and $h=\kappa L/D$. For each $n=1,2,\ldots$, $\alpha_n$ belongs
to an interval $[\pi(n-1),\pi(n-1/2))$ that facilitates their numerical
computation. Note that this problem is equivalent to diffusion on the interval
$(-L,L)$ whose both endpoints are partially reactive. The spectral expansion
of the survival probability is \cite{Grebenkov17e}
\begin{eqnarray}
\nonumber
S(t|x_0)&=&2\sum\limits_{n=1}^\infty\frac{(-1)^{n-1}h\sqrt{h^2+\alpha_n^2}}{
\alpha_n(h^2+h+\alpha_n^2)}\\
&&\times e^{-Dt\alpha_n^2/L^2}\cos(\alpha_nx_0/L).
\end{eqnarray}
Its volume average reads
\begin{equation}
\overline{S(t)}=2h^2\sum\limits_{n=1}^\infty\frac{e^{-Dt\alpha_n^2/L^2}}{
\alpha_n^2(h^2+h+\alpha_n^2)}.
\end{equation}
The probability density $\rho(t|x_0)$ and its volume average $\overline{
\rho(t)}$ follow by taking the time derivative.

\subsection{Absorbing sphere}

We also consider the first-passage problem to the absorbing boundary of a ball
of radius $L$, for which the survival probability is also explicitly known,
\begin{subequations}
\begin{eqnarray}
\nonumber
S(t|\x_0)&=&1-\frac{R}{|\x_0|}\sum\limits_{k=1}^\infty\biggl[\erf\biggl(\frac{
(2k-1)L+|\x_0|}{\sqrt{4Dt}}\biggr)\\  
&&-\erf\biggl(\frac{(2k-1)L-|\x_0|}{\sqrt{4Dt}}\biggr)\biggr]\\
\nonumber
&=&2\sum\limits_{n=1}^\infty(-1)^{n+1}\frac{\sin(\pi n|\x_0|/L)}{\pi n|\x_0|/L}\\
&&\times e^{-\pi^2n^2Dt/L^2}. 
\end{eqnarray}
\end{subequations}

As a consequence, the PDF reads
\begin{widetext}
\begin{eqnarray}
\nonumber
\rho(t|\x_0)&=&\frac{L}{|\x_0|\sqrt{4\pi Dt^3}}\sum\limits_{k=1}^\infty\biggl(
((2k-1)L-|\x_0|)e^{-((2k-1)L-|\x_0|)^2/(4Dt)}-((2k-1)L+|\x_0|)e^{-((2k-1)L+|
\x_0|)^2/(4Dt)}\biggr)\\
&=&\frac{2D}{L^2}\sum\limits_{n=1}^\infty(-1)^{n+1}\frac{\sin(\pi n|\x_0|/L)}{
\pi n|\x_0|/L}(\pi n)^2e^{-\pi^2n^2Dt/L^2}.
\end{eqnarray}
\end{widetext}

We also get
\begin{subequations}
\begin{eqnarray}   
\nonumber
\overline{S(t)}&=&1+\frac{3Dt}{L^2}-\frac{6\sqrt{Dt}}{L\sqrt{\pi}}\biggl\{1   
+2\sum\limits_{k=1}^\infty\biggl(e^{-k^2 L^2/(Dt)}\\
&&-\sqrt{\pi}(k L/\sqrt{Dt})\erfc\bigl(kL/\sqrt{Dt}\bigr)\biggr)\biggr\}\\
\label{eq:S3d_volume}
&=&\frac{6}{\pi^2}\sum\limits_{n=1}^\infty\frac{e^{-\pi^2 n^2 Dt/L^2}}{n^2} 
\end{eqnarray}
\end{subequations}
and
\begin{subequations}
\begin{eqnarray}
\nonumber
\overline{\rho(t)}&=&-\frac{3D}{L^2}+\frac{3\sqrt{D}}{L\sqrt{\pi t}}\\
&&+\frac{6\sqrt{D}}{L\sqrt{\pi t}}\sum\limits_{k=1}^\infty e^{-k^2L^2/(Dt)}\\
&=&\frac{6D}{L^2}\sum\limits_{n=1}^\infty e^{-\pi^2n^2Dt/L^2}.
\end{eqnarray}
\end{subequations}

We also easily get
\begin{equation}
\overline{\delta^2(\x_0)}=\frac{4\pi}{4\pi L^3/3}\int\limits_0^L drr^2(L-r)^2=
\frac{L^2}{10}.
\end{equation}
and
\begin{eqnarray}
\nonumber
b_1&=&\ln\biggl(\frac{1}{\sqrt{4\pi L^3/3}}\int\limits_{\Omega}d\x\beta_1\frac{
\sin(\pi|\x|/L)}{\pi|\x|/L}\biggr)\\  \nonumber
&&+\frac{1}{4\pi L^3/3}\int\limits_{\Omega}d\x\ln\biggl(\sqrt{4\pi L^3/3}\beta_1
\frac{\sin(\pi|\x|/L)}{\pi|\x|/L}\biggr)\\
&=&-\frac{3\zeta(3)}{2\pi^2}-2\ln\pi,
\end{eqnarray}
where $\zeta(z)$ is the Riemann zeta function with $\zeta(3)\approx1.2021$.
Here we used the following expression for the first eigenfunction: $u_1(\x)=
\beta_1\frac{\sin(\pi|\x|/L)}{\pi|\x|/L}$, where $\beta_1=\sqrt{\pi/(2L^3)}$
is the normalisation factor. Note also that $c_1=6/\pi^2$.

\subsection{Spherical target surrounded by a reflecting sphere}
\label{sec:targetS}

We also consider a more elaborate situation of a spherical target of
radius $R$, which is surrounded by a larger reflecting concentric
sphere of radius $L$.  The related first-passage time distribution for
a single particle was studied in \cite{Grebenkov18}.  In particular,
the exact spectral expansion for the survival probability was
provided:
\begin{equation}  \label{eq:Ht}
S(t|\x_0) = \sum\limits_{n=1}^\infty u_n(|\x_0|) \, e^{- Dt \lambda_n} ,
\end{equation}
where
\begin{widetext}
\begin{eqnarray*}
\lambda_n &=& \alpha_n^2/(L-R)^2 , \\  \label{eq:un}
u_n(r) &=& \frac{b_n}{\lambda_n (L-R)^2}
\frac{L\alpha_n\cos\bigl(\alpha_n\frac{L-r}{L-R}\bigr)-(L-R)\sin
\bigl(\alpha_n\frac{L-r}{L-R}\bigr)}{r\alpha_n}, \\
\label{cn}
b_n &=& \frac{2R^2\alpha_n^2}{(R L+\mu(L^2+R^2))\alpha_n\sin\alpha_n+R(
\mu L\alpha_n^2-R)\cos\alpha_n} \,,
\end{eqnarray*}
\end{widetext}
$\mu = D/(\kappa (L-R))$, and $\alpha_n$ (with $n = 1,2,\ldots$)
are positive solutions of the equation
\begin{equation}
\label{eq:alphan_sphere}
\tan\alpha_n=\frac{\alpha_n\bigl(R L+\mu(L-R)^2\bigr)}{R(L-R)+\mu(L-
R)^2+\mu LR\alpha_n^2} \,.
\end{equation}
One easily gets $H(t|\x_0)$ by taking the time derivative.

The volume-averaged survival probability is obtained by integration
over $\x_0$,
\begin{eqnarray} \nonumber
\overline{S(t)} &=& \frac{1}{\frac{4\pi}{3} (L^3-R^3)} \sum\limits_{n=1}^\infty e^{- Dt \lambda_n} 4\pi \int\limits_{R}^{L} dr \, r^2 \, u_n(r) \\
\label{eq:Sbar_target}
&=& \sum\limits_{n=1}^\infty c_n \, e^{-D\lambda_n t} \,,
\end{eqnarray}
with
\begin{equation}
\label{eq:cn_sphere}
c_n=\frac{3(L-R)^3}{L^3-R^3}\frac{b_n\bigl[\bigl(1+ \alpha_n^2 \frac{R L}{(L-R)^2}\bigr)\sin\alpha_n-
\alpha_n\cos\alpha_n\bigr]}{\alpha_n^5}  \,.
\end{equation}

\section{The closest starting point}
\label{sec:closest}

In this Appendix, we generalise and rationalise the scaling relation
$\overline{\langle \T_N\rangle} \propto 1/N^2$ in terms of the closest
particle to the target.  Let us start with an example of the interval
$(0,L)$ with the absorbing endpoint $0$ and reflecting endpoint $L$.
For $N$ independent particles randomly placed on the interval at
positions $x_1,\ldots,x_N$, the distance to the target at $0$ is
$\delta = \min\{x_1,\ldots,x_N\}$.  This is a random variable
characterised by the simple law: $\P\{\delta > x\} = (\P\{x_1 > x\})^N
= (1 - x/L)^N$, where we assumed the uniform distribution for $x_i$.
As a consequence, the mean shortest distance is
\begin{equation}
\overline{\delta} = \int\limits_0^L dx \, x \, \bigl(-\partial_x \P\{\delta > x\}\bigr) = 
\int\limits_0^L dx \,\P\{\delta > x\} = \frac{L}{N+1} \,.
\end{equation}
Similarly, the second, the third, etc. closest particle are located on
average at the distance $2L/(N+1)$, $3L/(N+1)$, etc.  This is
expected: if one had to place $N$ particles at equal distances between
any two neighbours (and from the endpoints), one would get their
positions at $kL/(N+1)$, with $k = 1,2,\ldots,N$.  If $N$ is large,
there are many particles that start at the distance of the order $L/N$
from the target, and there is high chance that one of these particles
reaches the target within the time of the order of
$\overline{\delta}^2/D$, in agreement with
Eq. (\ref{eq:MfFPT_1d_approx}), up to a numerical prefactor.

The above argument can be easily extended to arbitrary bounded domains
with a target having a smooth boundary $\Gamma$.  Here, we define
again the shortest distance to the target as $\delta = \min\{|\x_1 -
\Gamma|,\ldots,|\x_N - \Gamma|\}$.  This random variable is again
characterised by the law $\P\{\delta > x\} = (\P\{ |\x_1 - \Gamma| >
x\})^N$ due to the independence of the starting points.  Even though
the probability law for the distance from a single particle is now
complicated, its asymptotic behaviour remains simple.  Indeed, for
small $x$, $\P\{|\x_1 - \Gamma| < x\}$ is the probability of locating
$\x_1$ within a thin boundary layer of width $x$ near the target
$\Gamma$.  For smooth $\Gamma$ and small $x$, the volume of this layer
can be estimated as $x |\Gamma|$, where $|\Gamma|$ is the surface area
of $\Gamma$.  As a consequence, $\P\{|\x_1 - \Gamma| < x\} \simeq x
|\Gamma|/|\Omega|$ as $x\to 0$.  The mean shortest distance is then
\begin{eqnarray*}
\overline{\delta} &=& \int\limits_0^{L} dx \, x \, \bigl(-\partial_x \P\{\delta > x\}\bigr) = 
\int\limits_0^L dx \,\P\{\delta > x\} \\
&=& \int\limits_0^L dx \,(1 - \P\{|\x_1 - \Gamma| < x\})^N,
\end{eqnarray*}
where $L$ is the maximal distance from the target: $L =
\max_{\x\in\Omega} |\x - \Gamma|$.  For large $N$, the main
contribution comes from $x$ close $0$, for which $\P\{|\x_1 - \Gamma|
< x\}$ is small.  Substituting the above approximation in this region,
one gets
\begin{equation}  \label{eq:delta}
\overline{\delta} \simeq \int\limits_0^{L'} dx \,(1 - x |\Gamma|/|\Omega|)^N \simeq \frac{|\Omega|}{|\Gamma|(N+1)} \,,
\end{equation}
where the integral was truncated at $L' = |\Omega|/|\Gamma|$ to ensure
that the integrand function remains positive.  Similar computation can
be performed to estimate that the mean distance to the target from the
next-to-the-closest points is of the same order.  If
$\overline{\delta}$ is small as compared to the mean curvature of the
target, diffusion in the lateral direction does not matter, and the
problem is again reduced to diffusion of a single particle on the
interval $(0,\overline{\delta})$, for which the mean FPT is again
$\overline{\delta}^2/(2D)$.  This rationalises the scaling relation
$\overline{\langle \T_N\rangle} \propto 1/N^2$ that we derived in
Eq. (\ref{eq:MfFPT_1d_approx}).

An important consequence of this reasoning is that the scaling $1/N^2$
holds far beyond the considered case of the uniform distribution of
starting points.  First, one can realise that only the behaviour of the
distribution near the boundary of the target matters in the limit of
large $N$; indeed, the integrand function $(1 - \P\{|\x_1 - \Gamma| <
x\})^N$ becomes exponentially small when $\P\{|\x_1 - \Gamma| < x\}$
is not small.  Second, even in the vicinity of the target, the
distribution does not need to be uniform.  For instance, if $\P\{|\x_1
- \Gamma| < x\} \simeq (x/A)^\alpha$ with some scale $A> 0$ and any
exponent $\alpha > 0$, the integral in Eq. (\ref{eq:delta}) yields
$\overline{\delta} \simeq A/(\alpha N + 1)$, i.e., the scaling $1/N$
is just affected by a prefactor.  This result confirms the generality
of the scaling relation (\ref{eq:MfFPT_1d_approx}).  This conclusion
can also be obtained on a more rigorous basis by considering the heat
kernel short-time asymptotic behaviour.  In fact, the asymptotic
relation (\ref{eq:Svol_short}) for the volume-averaged survival
probability holds in a much more general setting, in particular, after
the average with a smooth nonuniform distribution of the starting
point \cite{vandenBerg89,Desjardins94,Gilkey}.  We do not explore this
direction in the paper.

Only if the distribution of the starting point excludes points near
the target, the volume-averaged mean fFPT would scale differently.  A
standard example is the Dirac distribution that fixes the starting
point, which was studied previously and yielded the $\overline{\langle
\T_N\rangle} \propto 1/\ln N$ behaviour \cite{Weiss83} (see also
Sec. \ref{sec:intro} for more references). A similar behaviour is
expected for any distribution that excludes points from some $\delta_0$-vicinity
of the target, i.e., if there exists $\delta_0 > 0$ such that
$\P\{ |\x_1 - \Gamma| < \delta_0\} = 0$.  Moreover, this strict
exclusion condition can be relaxed.  For instance, if $\P\{ |\x_1 -
\Gamma| < x\} \simeq e^{-A/x}$ as $x\to 0$ (with some scale $A > 0$),
the integral in Eq. (\ref{eq:delta}) yields again the $1/\ln N$
behaviour for large $N$.  A more systematic analysis of this problem
and finding the sharp exclusion condition, present interesting
problems for future research.


\begin{thebibliography}{99}

\bibitem{atkins} P. Atkins, J. de Paula, and J. Keeler, Physical chemistry (Oxford
University Press, Oxford, UK, 2017)

\bibitem{smol} M. v. Smoluchowski, Drei Vortr\"age \"uber Diffusion, Brownsche
Molekularbewegung und Koagulation von Kolloidteilchen, Z. Phys. {\bf 17}, 557
(1916).

\bibitem{Collins49} F. C. Collins and G. E. Kimball, Diffusion-controlled
reaction rates, J. Coll. Sci. {\bf 4}, 425 (1949).

\bibitem{bvh} P. H. von Hippel and O. G. Berg, Facilitated Target Location in
Biological Systems, J. Biol. Chem. \textbf{264}, 675 (1989).

\bibitem{leonid} M. Slutsky and L. A. Mirny, Kinetics of Protein-DNA Interaction:
Facilitated Target Location in Sequence-Dependent Potential, Biophys. J. \textbf{87},
4021 (2004).

\bibitem{olivier} M. Coppey, O. B{\'e}nichou, R. Voituriez, and M. Moreau,
Kinetics of Target Site Localization of a Protein on DNA: A Stochastic Approach,
Biophys. J. \textbf{87}, 1640 (2004).

\bibitem{michael} M. A. Lomholt, B. v. d. Broek, S.-M. J. Kalisch, G. J. L. Wuite,
and R. Metzler, Facilitated diffusion with DNA coiling, Proc. Natl. Acad. Sci. USA
\textbf{106}, 8204 (2009).

\bibitem{gijs} B. van den Broek, M. A. Lomholt, S.-M. J. Kalisch, R. Metzler,
and G. J. L. Wuite, How DNA coiling enhances target localization by proteins,
Proc. Natl. Acad. Sci. USA \textbf{105}, 15738 (2008).

\bibitem{austin} Y. M. Wang, R. H. Austin, and E. C. Cox, Single Molecule
Measurements of Repressor Protein 1D Diffusion on DNA, Phys. Rev. Lett.
\textbf{97}, 048302 (2006).

\bibitem{golding} I. Golding, J. Paulsson, S. M. Zawilski, and E. C. Cox,
Real-Time Kinetics of Gene Activity in Individual Bacteria, Cell \textbf{123},
1025 (2005).

\bibitem{xie} J. Elf, G.-W. Li, and X. S. Xie, Probing Transctiption Factor
Dynamics at the Single Molecule Level a Living Cell, Science \textbf{316}, 1191
(2007).

\bibitem{mark} I. M. Sokolov, R. Metzler, K. Pant, and M. C. Williams, Target
search of $N$ sliding proteins on a DNA, Biophys. J. \textbf{89}, 895 (2005).

\bibitem{elf} D. L. Jones, P. Leroy, C. Unoson, D. Fange, V. Curic, M. J.
Lawson, and J. Elf, Kinetics of dCas9 target search in Escherichia coli,
Science \textbf{357}, 1420 (2017).

\bibitem{elf1} P. Hammar, P. Leroy, A. Mahmutovic, E. G. Marklund, O. G. Berg,
and J. Elf, The lac repressor displays facilitated diffusion in living cells,
Science \textbf{336}, 1595 (2012).

\bibitem{calef} D. Calef and J. M. Deutch, Diffusion-Controlled Reactions,
Annu. Rev. Phys. Chem. {\bf 34}, 493 (1983).

\bibitem{szabo} A. Szabo, K. Schulten, and Z. Schulten, First passage time
approach to diffusion controlled reactions, J. Chem. Phys. {\bf 72}, 4350 (1980).

\bibitem{weiss} G. H. Weiss, Overview of theoretical models for reaction rates,
J. Stat. Phys. {\bf 42}, 3 (1986).

\bibitem{Hanggi90} P. H\"anggi, P. Talkner, and M. Borkovec, Reaction-rate theory:
fifty years after Kramers, Rev. Mod. Phys. {\bf 62}, 251 (1990).

\bibitem{szabo2} A. Szabo, Theoretical approaches to reversible diffusion-influenced
reactions: Monomer-excimer kinetics, J. Chem. Phys. {\bf 95},  2481 (1991).

\bibitem{colloids} G. Oshanin, M. Moreau, and S. F. Burlatsky, Models of chemical
reactions with participation of polymers, Adv. Colloid \& Interface Sci. {\bf 49},
1 (1994).

\bibitem{ralf} E. Gudowska-Nowak, K. Lindenberg, and R. Metzler, Preface: Special
issue on Marian Smoluchowski's $1916$ paper -- a century of inspiration, 
J. Phys. A: Math. Theor. {\bf 50},  380301 (2017).

\bibitem{Benichou14} O. B\'enichou and R. Voituriez, From first-passage times of
random walks in confinement to geometry-controlled kinetics, Phys. Rep. {\bf 539},
225-284 (2014).
                            
\bibitem{Oshanin} K. Lindenberg, R. Metzler, and G. Oshanin (Eds.), Chemical
Kinetics: Beyond the Textbook (New Jersey: World Scientific, 2019). 
				
\bibitem{Grebenkov19b} D. S. Grebenkov, Imperfect Diffusion-Controlled Reactions, 
in Chemical Kinetics: Beyond the Textbook, Eds. K. Lindenberg, R. Metzler, and
G. Oshanin (World Scientific, 2019; available online as ArXiv: 1806.11471).

\bibitem{bal} B. Ya. Balagurov and V. G. Vaks, Random walks of a particle on
lattices with traps, Sov. Phys. JETP {\bf 38}, 968 (1974).
 
\bibitem{don} M. D. Donsker and S. R. S. Varadhan, Asymptotics for the wiener
sausage, Commun. Pure Appl. Math. {\bf 28}, 525 (1975).

\bibitem{bur} S. F. Burlatsky, The influence of spatial non-homogeneity on the
kinetics of bimolecular reactions, Theor. Exp. Chem. {\bf 14}, 373 (1979).

\bibitem{ovch1} A. A. Ovchinnikov and Ya. B. Zeldovich, Role of density
fluctuations in bimolecular reaction kinetics, Chem. Phys. {\bf 28}, 215 (1978).

\bibitem{red} K. Kang and S. Redner, Scaling Approach for the Kinetics of
Recombination Processes, Phys. Rev. Lett. {\bf 52}, 955 (1984).

\bibitem{ovch2} G. Oshanin, A. A. Ovchinnikov, and S. F. Burlatsky,
Fluctuation-induced kinetics of reversible reactions, J. Phys. A: Math.
Gen. {\bf 22}, L977 (1989); Fluctuation-induced kinetics of reversible
coagulation, J. Phys. A: Math. Gen. {\bf 22}, L973  (1989); S. F. Burlatskii,  
A. A. Ovchinnikov, and G. Oshanin, Fluctuation kinetics in systems with 
reversible recombination, JETP {\bf 68}, 1153 (1989) [Zh. Eksp. Teor. Fiz. {\bf 95},1993 (1989)] 

\bibitem{bray} A. J. Bray and R. A. Blythe, Exact Asymptotics for One-Dimensional
Diffusion with Mobile Traps, Phys. Rev. Lett. {\bf 89}, 150601 (2002); 
R. A. Blythe and A. J. Bray, Survival probability of a diffusing particle in the
presence of Poisson-distributed mobile traps, Phys. Rev. E {\bf 67}, 041101 (2003).

\bibitem{osh} G. Oshanin, O. B\'enichou, M. Coppey, and M. Moreau, Trapping
reactions with randomly moving traps: Exact asymptotic results for compact
exploration, Phys. Rev. E {\bf 66}, 060101(R) (2002).

\bibitem{osh2} S. B. Yuste, G. Oshanin, K. Lindenberg, O. B\'enichou, and J. Klafter, 
Survival probability of a particle in a sea of mobile traps: A tale of tails,
Phys. Rev. E {\bf 78}, 021105 (2008).
 
\bibitem{tauber} U. C. T\"auber, Fluctuations and correlations in chemical reaction
kinetics and population dynamics, in:  Chemical Kinetics Beyond the Textbook, K.
Lindenberg, R. Metzler, and G. Oshanin (Eds.) (World Scientific, Singapore, 2019).

\bibitem{conc1} B. U. Felderhof and J. M. Deutch, Concentration dependence of the
rate of diffusion-controlled reactions, J. Chem. Phys. {\bf 64}, 4551 (1976).

\bibitem{conc2} S. F. Burlatsky, G. Oshanin, and A. A. Ovchinnikov,  
Kinetics of chemical short-range ordering in liquids and diffusion-controlled
reactions, Chem. Phys. {\bf 152}, 13 (1991).

\bibitem{conc3} K. Seki and M. Tachiya, Reaction under vacancy-assisted diffusion
at high quencher concentration, Phys. Rev. E {\bf 80}, 041120 (2009).

\bibitem{conc4} M. Kim, S. Lee, and J-H. Kim, Concentration effects on the rates
of irreversible diffusion-influenced reactions, J. Chem. Phys. {\bf 141}, 084101
(2014).

\bibitem{alberts} B. Alberts et al., Molecular Biology of the Cell (Garland
Science, New York, NY, 2014).

\bibitem{snustad} D. P. Snustad and M. J. Simmons, Principles of Genetics 
(Wiley, New York, 2000).

\bibitem{kuhlman} T. E. Kuhlman and E. C. Cox, Gene location and DNA
density determine transcription factor distributions in Escherichia coli,
Mol. Syst. Biol. {\bf 8}, 610 (2012).

\bibitem{kepes} F. K{\'e}p{\`e}s, Periodic Transcriptional Organization of the
E. coli Genome, J. Mol. Biol. \textbf{340}, 957 (2004).

\bibitem{kolesov} G. Kolesov, Z. Wunderlich, O. N. Laikova, M. S. Gelfand, and
L. A. Mirny, Proc. Natl. Acad. Sci. USA \textbf{104}, 13948 (2007).

\bibitem{pulkkinen} O. Pulkkinen and R. Metzler, Distance matters: the impact of
gene proximity in bacterial gene regulation, Phys. Rev. Lett. \textbf{110},
198101 (2013).

\bibitem{prathit} P. Kar, A. G. Cherstvy, and R. Metzler, Acceleration of bursty
multi-protein target-search kinetics on DNA by colocalisation, Phys. Chem. Chem.
Phys. \textbf{20}, 7931 (2018).

\bibitem{mcadams} H. H. McAdams and A. Arkin, It's a noisy business! Genetic
regulation at the nanomolar scale, Trends in Genet. \textbf{15}, 65 (1999).

\bibitem{mcadams1} H. H. McAdams and A. Arkin, Stochastic mechanisms in gene
expression, Proc. Natl. Acad. Sci. USA \textbf{94}, 814 (1997).

\bibitem{kuttler} J. M{\"u}ller, C. Kuttler, and B. A. Hense, Sensitivity of the
quorum sensing system is achieved by low pass filtering, Biosystems \textbf{92},
76 (2008).

\bibitem{mugler} S. Fancher and A. Mugler, Fundamental limits to collective
concentration sensing in cell populations, Phys. Rev. Lett. \textbf{118}, 078101
(2017).

\bibitem{seno} M. Marenda, M. Zanardo, A. Trovato, F. Seno, and A. Squartini,
Modeling quorum sensing trade-offs between bacterial cell density and system
extension from open boundaries, \textit{Sci. Rep.} \textbf{6}, 39142 (2016).

\bibitem{trovato} A. Trovato, F. Seno, M. Zanardo, S. Alberghini, A. Tondello,
and A. Squartini, Quorum vs. diffusion sensing: a quantitative analysis of the
relevance of absorbing or reflecting boundaries, FEMS Microbiol. Lett.
\textbf{352}, 198 (2014).

\bibitem{oliver} O. Kindler, O. Pulkkinen, A. Cherstvy, and R. Metzler, Burst
statistics in an early biofilm quorum sensing model: the role of spatial
colony-growth heterogeneity, Sci. Rep. \textbf{9}, 12077 (2019).

\bibitem{oudenaarden} E. M. Ozbudak, M. Thattai, I. Kurtser, A. D. Grossman, and
A. van Oudenaarden, Regulation of noise in the expression of a single gene,
Nature Genet. \textbf{31}, 69 (2002).

\bibitem{oberg} A. Mahmutovic, D. Fange, O. Berg, and J. Elf,
Lost in presumption: stochastic reactions in spatial models,
Nature Meth. \textbf{9}, 1163 (2012).

\bibitem{carmine} C. Di Rienzo, E. Gratton, F. Beltram, and F. Cardarelli,
Spatiotemporal Fluctuation Analysis: A Powerful Tool for the Future Nanoscopy
of Molecular Processes, Biophys. J. \textbf{111}, 679 (2016).

\bibitem{Grebenkov18o}  D. Grebenkov, R. Metzler, and G. Oshanin, Towards a full
quantitative description of single-molecule reaction kinetics in biological
cells, Phys. Chem. Chem. Phys. {\bf 20}, 16393 (2018).

\bibitem{Grebenkov18} D. S. Grebenkov, R. Metzler, and G. Oshanin, Strong
defocusing of molecular reaction times results from an interplay of geometry
and reaction control, Commun. Chem. {\bf 1}, 96 (2018).

\bibitem{GrebenkovFPT} D. S. Grebenkov, R. Metzler, and G. Oshanin, 
Full distribution of first exit times in the narrow escape problem, 
New J. Phys. {\bf 21}, 122001 (2019).

\bibitem{Sano79}  H. Sano and M. Tachiya, Partially diffusion-controlled
recombination, J. Chem. Phys. {\bf 71}, 1276 (1979).

\bibitem{Sapoval94} B. Sapoval, General Formulation of Laplacian Transfer
Across Irregular Surfaces, Phys. Rev. Lett. {\bf 73}, 3314 (1994).

\bibitem{Grebenkov06} D. S. Grebenkov, Partially Reflected Brownian Motion:
A Stochastic Approach to Transport Phenomena, in Focus on Probability Theory,
Ed. L. R. Velle, pp. 135-169 (Nova Science Publishers, 2006). 

\bibitem{Grebenkov07a} D. S. Grebenkov, Residence times and other functionals
of reflected Brownian motion, Phys. Rev. E {\bf 76}, 041139 (2007). 

\bibitem{Singer08} A. Singer, Z. Schuss, A. Osipov, and D. Holcman,
Partially reflected diffusion, SIAM J. Appl. Math. {\bf 68}, 844-868 (2008).

\bibitem{Bressloff08} P. C. Bressloff, B. A. Earnshaw, and M. J. Ward,
Diffusion of protein receptors on a cylindrical dendritic membrane with partially
absorbing traps, SIAM J. Appl. Math. {\bf 68}, 1223 (2008).

\bibitem{Grebenkov10b} D. S. Grebenkov, Subdiffusion in a bounded domain with a
partially absorbing-reflecting boundary, Phys. Rev. E {\bf 81}, 021128 (2010).

\bibitem{Grebenkov19} D. S. Grebenkov, Spectral theory of imperfect
diffusion-controlled reactions on heterogeneous catalytic surfaces,
J. Chem. Phys. {\bf 151}, 104108 (2019). 

\bibitem{Grebenkov20} D. S. Grebenkov,
Paradigm shift in diffusion-mediated surface phenomena,
Phys. Rev. Lett. {\bf 125}, 078102 (2020).

\bibitem{thiago} T. Mattos, C. Mej\'{\i}a-Monasterio, R. Metzler, and G. Oshanin,
First passages in bounded domains: When is the mean first passage
time meaningful?, Phys. Rev. E \textbf{86}, 031143 (2012).

\bibitem{Rupprecht15} J.-F. Rupprecht, O. B\'enichou, D. S. Grebenkov, and R.
Voituriez, Exit time distribution in spherically symmetric two-dimensional
domains, J. Stat. Phys. {\bf 158}, 192-230 (2015).

\bibitem{aljaz} A. Godec and R. Metzler, Universal proximity effect in target search
kinetics in the few encounter limit, Phys. Rev. X \textbf{6}, 041037 (2016).

\bibitem{aljaz1}  A. Godec and R. Metzler, First passage time distribution in
heterogeneity controlled kinetics: going beyond the mean first passage time,
Sci. Rep. \textbf{6}, 20349 (2016).

\bibitem{biofilm} M. Klinger-Strobel, H. Suesse, D. Fischer, M. W. Pletz, and
O. Makarewicz, A Novel Computerized Cell Count Algorithm for Biofilm Analysis,
PLoS ONE \textbf{11}, e0154937 (2016).

\bibitem{biofilm1} H.-C. Flemming and S. Wuertz, Bacteria and archaea on Earth and
their abundance in biofilms, Nature Rev. Microbiol. \textbf{17}, 247 (2019).

\bibitem{holcman} Z. Schuss, K. Basnayake, and D. Holcman, Redundancy principle
and the role of extreme statistics in molecular and cellular biology,
Phys. Life Rev. {\bf 28}, 52 (2019). 

\bibitem{Reynaud15} K. Reynaud, Z. Schuss, N. Rouach, and D. Holcman,
Why so many sperm cells?, Commun. Integr. Biol. {\bf 8}, e1017156 (2015).
					
\bibitem{Basnayake19} K. Basnayake, Z. Schuss, and D. Holcman, Asymptotic
formulas for extreme statistics of escape times in 1, 2 and 3-dimensions,
J. Nonlinear Sci. {\bf 29}, 461-499 (2019).

\bibitem{Phillips} R. Phillips and R. Milo, Cell Biology by the Numbers
(Garland Sciences, NY, 2015).

\bibitem{fain} G. L. Fain, Molecular and cellular physiology of neurons, 
(Harvard University Press, 1999)

\bibitem{Weiss83} G. H. Weiss, K. E. Shuler, and K. Lindenberg, Order Statistics
for First Passage Times in Diffusion Processes, J. Stat. Phys. {\bf 31}, 255-278
(1983).

\bibitem{Abad12} E. Abad, S. B. Yuste, and K. Lindenberg, Survival probability of
an immobile target in a sea of evanescent diffusive or subdiffusive traps: A
fractional equation approach, Phys. Rev. E {\bf 86}, 061120 (2012).
					
\bibitem{Meerson15} B. Meerson and S. Redner, Mortality, redundancy, and diversity
in stochastic search, Phys. Rev. Lett. {\bf 114}, 198101 (2015).
					
\bibitem{Meerson15a} B. Meerson, The number statistics and optimal history of
non-equilibrium steady states of mortal diffusing particles, J. Stat. Mech.
P05004 (2015).
                        
\bibitem{Lawley19} S. D. Lawley and J. B. Madrid, A probabilistic approach to
extreme statistics of Brownian escape times in dimensions 1, 2, and 3,
J. Nonlin. Sci. {\bf 30}, 1207-1227 (2020).

\bibitem{Lawley19b} S. D. Lawley, Universal Formula for Extreme First Passage
Statistics of Diffusion, Phys. Rev. E {\bf 101}, 012413 (2020).

\bibitem{Lawley19c} S. D. Lawley, Distribution of extreme first passage times
of diffusion, J. Math. Biol. {\bf 80}, 2301-2325 (2020).


\bibitem{Ro17} S. Ro and Y. W. Kim, Parallel random target searches in a confined space,
Phys. Rev. E {\bf 96}, 012143 (2017).

\bibitem{Agranov18} T. Agranov and B. Meerson, Narrow Escape of Interacting Diffusing Particles,
Phys. Rev. Lett. {\bf 120}, 120601 (2018).

\bibitem{Madrid20} J. Madrid and S. D. Lawley, 
Competition between slow and fast regimes forextreme first passage
times of diffusion, J. Phys. A: Math. Theor. {\bf 53}, 335002 (2020).





\bibitem{mejia} C. Mej'{i}a-Monasterio, G. Oshanin, and G.  Schehr, First passages
for a search by a swarm of independent random searchers,
J. Stat. Mech. P06022 (2011).

\bibitem{mejia2} T. G Mattos, C. Mej\'{i}a-Monasterio, R. Metzler, G.  Oshanin, and G. Schehr, 
Trajectory-to-trajectory fluctuations in first-passage phenomena in bounded domains, in: 
First-passage phenomena and their applications, eds. R. Metzler, G. Oshanin and S. Redner
(World Scientific Publ., Singapore, 2014), pp. 203-225

\bibitem{Bray13}  A. J. Bray, S. Majumdar, and G. Schehr,
Persistence and First-Passage Properties in Non-equilibrium Systems,
Adv. Phys. {\bf 62}, 225-361 (2013).

\bibitem{tachiya} M. Tachiya, Theory of diffusion-controlled reactions:
Formulation of the bulk reaction rate in terms of the pair probability, 
Radial. Phys. Chem. {\bf 21}, 167 (1983). 

\bibitem{blumen} A. Blumen, G. Zumofen, and J. Klafter, Target annihilation
by random walkers, Phys. Rev. B {\bf 30}, 5379(R) (1984).

\bibitem{sergei} S. F. Burlatsky and A. A. Ovchinnikov, Effect of
reactant-fluctuation density on the kinetics of recombination, multiplication,
and trapping processes, Zh. Eksp. Teor. Fiz. 1618 (1987) [Sov. Phys. JETP
\textbf{65}, 908 (1987)].

\bibitem{agmon}  A. Szabo, R. Zwanzig, and N. Agmon, Diffusion-Controlled
Reactions with Mobile Traps, Phys. Rev. Lett. {\bf 61}, 2496 (1988).

\bibitem{searchjpa} R. Metzler, T. Koren, B. v. d. Broek, G. J. L. Wuite, and
M. A. Lomholt, And did he search for you, and could not find you?, J. Phys. A
\textbf{42}, 434005 (2009).


\bibitem{Benichou00} O. B\'enichou, M. Moreau, and G. Oshanin, Kinetics of
stochastically gated diffusion-limited reactions and geometry of random walk
trajectories, Phys. Rev. E {\bf 61}, 3388 (2000).
					
\bibitem{Lawleynon} S. D. Lawley and C. E. Miles, Diffusive Search for Diffusing
Targets with Fluctuating Diffusivity and Gating, J. Nonlinear Sci. {\bf 29},
2955 (2019).




\bibitem{Redner} S. Redner, A Guide to First Passage Processes (Cambridge:
Cambridge University Press, 2001).



\bibitem{Grigoriev02} I. V. Grigoriev, Y. A. Makhnovskii, A. M. Berezhkovskii,
and V. Y. Zitserman, Kinetics of escape through a small hole, J. Chem. Phys.
{\bf 116}, 9574-9577 (2002).

\bibitem{Holcman04} D. Holcman and Z. Schuss, Escape Through a Small Opening:
Receptor Trafficking in a Synaptic Membrane, J. Stat. Phys. {\bf 117}, 975-1014
(2004).

\bibitem{Schuss07} Z. Schuss, A. Singer, and D. Holcman, The narrow escape
problem for diffusion in cellular microdomains, Proc. Nat. Acad. Sci. USA
{\bf 104}, 16098 (2007).

\bibitem{Benichou08} O. B\'enichou and R. Voituriez, Narrow-Escape Time Problem:
Time Needed for a Particle to Exit a Confining Domain through a Small Window,
Phys. Rev. Lett. {\bf 100}, 168105 (2008).


\bibitem{Pillay10} S. Pillay, M. J. Ward, A. Peirce, and T. Kolokolnikov,
An Asymptotic Analysis of the Mean First Passage Time for Narrow Escape Problems:
Part I: Two-Dimensional Domains, SIAM Multi. Model. Simul. {\bf 8}, 803-835 (2010).

\bibitem{Cheviakov10} A. F. Cheviakov, M. J. Ward, and R. Straube, An Asymptotic
Analysis of the Mean First Passage Time for Narrow Escape Problems: Part II: The
Sphere, SIAM Multi. Model. Simul. {\bf 8}, 836-870 (2010).

\bibitem{Oshanin10} G. Oshanin, M. Tamm, and O. Vasilyev, Narrow-escape times for diffusion in microdomains 
with a particle-surface affinity: Mean-field results, J. Chem. Phys. {\bf 132}, 235101 (2010).

\bibitem{Cheviakov12} A. F. Cheviakov, A. S. Reimer, and M. J. Ward, Mathematical
modeling and numerical computation of narrow escape problems, Phys. Rev. E {\bf
85}, 021131 (2012).


\bibitem{Grebenkov16}  D. S. Grebenkov, 
Universal formula for the mean first passage time in planar domains, 
Phys. Rev. Lett. {\bf 117}, 260201 (2016)

\bibitem{GrebenkovNEP} D. S. Grebenkov and G. Oshanin,
Diffusive escape through a narrow opening: new insights into a classic problem,
Phys. Chem.   Chem. Phys. {\bf 19}, 2723 (2017).

\bibitem{Holcman14} D. Holcman and Z. Schuss, The Narrow Escape Problem,
SIAM Rev. {\bf 56}, 213-257 (2014).


\bibitem{Gardiner} C. W. Gardiner, Handbook of stochastic methods for physics,
chemistry and the natural sciences (Springer: Berlin, 1985).

\bibitem{Risken} H. Risken, The Fokker-Planck equation (Springer, Berlin, 1989).


\bibitem{Grebenkov13}	D. S. Grebenkov and B.-T. Nguyen, 
Geometrical structure of Laplacian eigenfunctions, SIAM Rev. {\bf 55}, 601-667 (2013).


\bibitem{vandenBerg89}	M. van den Berg and E. B. Davies,
			Heat Flow out of Regions in Rm,
			Math. Z. {\bf 202}, 463 (1989).

\bibitem{vandenBerg94}	M. van den Berg and P. Gilkey,
			Heat Content Asymptotics of a Riemannian Manifold with Boundary,
			J. Func. Anal. {\bf 120}, 48 (1994).

\bibitem{Desjardins94} S. Desjardins and P. Gilkey, Heat content asymptotics
for operators of Laplace type with Neumann boundary conditions,
Math. Z. {\bf 215}, 251-268 (1994).

\bibitem{Gilkey}	P. Gilkey, Asymptotic Formulae in Spectral Geometry (Chapman and Hall/CRC, BocaRaton, FL, 2004).

\bibitem{Varadhan67}	S. R. S. Varadhan, 
On the Behavior of the Fundamental Solution of the Heat Equation with Variable Coefficients,
Comm. Pure Appl. Math. {\bf 20}, 431-455 (1967).

\bibitem{Grebenkov17d} D. S. Grebenkov and J.-F. Rupprecht, The escape problem
for mortal walkers, J. Chem. Phys. {\bf 146}, 084106 (2017).

\bibitem{Grebenkov10a} D. S. Grebenkov, Searching for partially reactive sites:
Analytical results for spherical targets, J. Chem. Phys. {\bf 132}, 034104 (2010).

\bibitem{Ward93} M. J. Ward and J. B. Keller, Strong Localized Perturbations of
Eigenvalue Problems, SIAM J. Appl. Math. {\bf 53}, 770-798 (1993).

\bibitem{Evans11} M. R. Evans and S. N. Majumdar, Diffusion with Stochastic
Resetting, Phys. Rev. Lett. {\bf 106}, 160601 (2011).

\bibitem{lubensky} T. C. Lubensky, Fluctuations in random walks with random
traps, Phys. Rev. A {\bf 30}, 2657 (1984).

\bibitem{renn} S. R. Renn, Statistics of random walks on trapped lattices, 
Nucl. Phys. B {\bf 275}, 273  (1986).

\bibitem{monthus} C. Monthus, G. Oshanin, A. Comtet, and S. F. Burlatsky,
Sample-size dependence of the ground-state energy
 in a one-dimensional localization problem,
Phys. Rev. E {\bf 54}, 231 (1996). 

\bibitem{Grebenkov17e}   D. S. Grebenkov, First passage times for multiple particles
with reversible target-binding kinetics, J. Chem. Phys. {\bf 147}, 134112 (2017).



\end{thebibliography}
\end{document}